\pdfoutput=1
\documentclass[10pt,twocolumn]{article}
\usepackage[a4paper,top=2.0cm,bottom=2.2cm,left=1.7cm,right=1.7cm,columnsep=6.5mm]{geometry}
\usepackage[T1]{fontenc}
\usepackage[helvratio=0.92]{newtxtext}
\usepackage{newtxmath}
\usepackage{amsmath,graphicx,booktabs,textcomp,microtype,xcolor}
\usepackage{siunitx}
\usepackage{algorithm}
\usepackage[noend]{algpseudocode}
\definecolor{accent}{HTML}{0B4F8A}
\definecolor{absbox}{gray}{0.93}
\usepackage[super,sort&compress,comma]{natbib}
\usepackage[colorlinks=true,allcolors=accent]{hyperref}
\usepackage{etoolbox}
\usepackage{caption}
\DeclareCaptionLabelSeparator{pipe}{ $|$ }
\AtBeginEnvironment{tabular}{\sffamily\fontsize{7.5}{10}\selectfont}
\usepackage[nobottomtitles*]{titlesec}

\titleformat{\section}{\sffamily\bfseries\large\color{accent}}{\thesection}{0.55em}{}
\titleformat{\subsection}{\sffamily\bfseries\normalsize\color{accent}}{\thesubsection}{0.5em}{}
\titlespacing*{\section}{0pt}{11pt}{4pt}
\titlespacing*{\subsection}{0pt}{8pt}{2pt}
\usepackage{fancyhdr}
\fancypagestyle{firstpage}{\fancyhf{}\fancyfoot[L]{\footnotesize\sffamily Dash and Arvanitis}\fancyfoot[R]{\footnotesize\sffamily\thepage}}
\usepackage{needspace}


\newcommand{\addtabletext}[1]{\par\medskip{\setlength{\leftskip}{9pt}\fontsize{7.5}{9.5}\selectfont #1}}
\newcommand{\movie}[1]{\par\medskip\noindent #1\par}

\graphicspath{{figures/}}

\begin{document}
\twocolumn[{%
\begin{flushleft}
{\sffamily\footnotesize\color{accent}\textbf{ARTICLE}\enspace$\cdot$\enspace PREPRINT\par}
\vspace{6pt}
{\sffamily\bfseries\fontsize{19}{23}\selectfont Differentiable topology optimization for high-fidelity acoustic holography beyond the thin-element approximation\par}
\vspace{9pt}
{\large Pradosh P. Dash$^{1}$ and Costas D. Arvanitis$^{1,2,\ast}$\par}
\vspace{4pt}
{\footnotesize $^{1}$George W. Woodruff School of Mechanical Engineering, Georgia Institute of Technology, Atlanta, GA 30332, USA. $^{2}$Wallace H. Coulter Department of Biomedical Engineering, Georgia Institute of Technology and Emory University, Atlanta, GA 30332, USA. $^{\ast}$Correspondence: costas.arvanitis@gatech.edu\par}
\vspace{6pt}
{\color{accent}\rule{\textwidth}{1.2pt}}
\end{flushleft}
\vspace{2pt}
\noindent\colorbox{absbox}{\parbox{\dimexpr\textwidth-2\fboxsep\relax}{\vspace{5pt}\textbf{Acoustic holography in the megahertz range can impact applications from manufacturing and non-destructive testing to transcranial ultrasound. Designing lens topologies for complex holograms in the sub-megahertz to low-megahertz regime most relevant to transcranial therapy is difficult, because wave propagation through an acoustically thick lens cannot be ignored. Here we show that the heterogeneous angular spectrum approach (HASA), which admits in-plane varying speed-of-sound maps and supports rapid differentiable optimization with the ADAM optimizer, can generate three-dimensional lens topologies directly. With it we quantify the failure of the thin-element approximation (TEA) at these frequencies, where lenses enter an acoustically thick regime ($L \gg \lambda$) governed by refractive walk-off, and we introduce a dimensionless refractive walk-off number $N_\mathrm{w}$ that defines a frequency-dependent operating regime. Topology optimization outperformed TEA-based design in all 45 configurations tested, with the largest structural gains at or below 1 MHz; for high-numerical-aperture designs the fidelity at 1 MHz matched that at 2 MHz, while structural similarity was highest at 0.5 MHz. The sub-megahertz to low-megahertz window is therefore a favourable operating regime for high-fidelity transcranial holography rather than one to be avoided. Free-field hydrophone measurements of 3D-printed lenses reproduced the designed patterns, and because the framework also accounts for aberration in heterogeneous media, patient-specific lenses corrected skull-induced aberration in simulation and ex vivo experiment, generalized across skull segments, and shaped volumetric beams through the human skull. Together these results provide a framework for designing simple, economical and high-performance ultrasound systems.}\vspace{5pt}}}
\par\vspace{5pt}
{\sffamily\footnotesize\color{accent}\textbf{Keywords:}\color{black}\ acoustic holography, topology optimization, transcranial ultrasound, thin-element approximation, angular spectrum method\par}
\vspace{12pt}
}]
\thispagestyle{firstpage}

\section{Introduction}
While therapeutic ultrasound has evolved significantly, its noninvasive application in the central nervous system remains constrained by the human skull, an acoustic barrier that attenuates and aberrates high-frequency waves \cite{R1}. Dense phased arrays can compensate for these aberrations and enable targeted neuro-interventions, but they are prohibitively expensive and their spatial resolution is fundamentally limited by electronic packaging constraints \cite{R2,R3}. Acoustic holography based on 3D-printed lenses offers an economical alternative for designing flexible ultrasound systems \cite{R4}. By encoding spatial phase patterns, these passive holograms turn a single-element transducer into a device that generates complex volumetric pressure fields for applications including contactless manufacturing \cite{R5}, imaging \cite{R6} and targeted neuro-interventions \cite{R7,R8}. Because their apparent element size is much smaller than the wavelength, large-aperture holograms provide reconstruction degrees of freedom equivalent to phased arrays that cannot be built ($10^{4}$ to $10^{5}$ elements) \cite{R4}. Numerical implementations that propagate waves in heterogeneous media can also pre-compensate for patient-specific skull aberration, producing precise focal spots for high-intensity focused ultrasound \cite{R7,R9} or targeting multiple brain regions at once for therapies such as blood-brain barrier opening \cite{R8,R10,R11}.

\begin{figure*}[t]
\centering
\includegraphics[width=\textwidth]{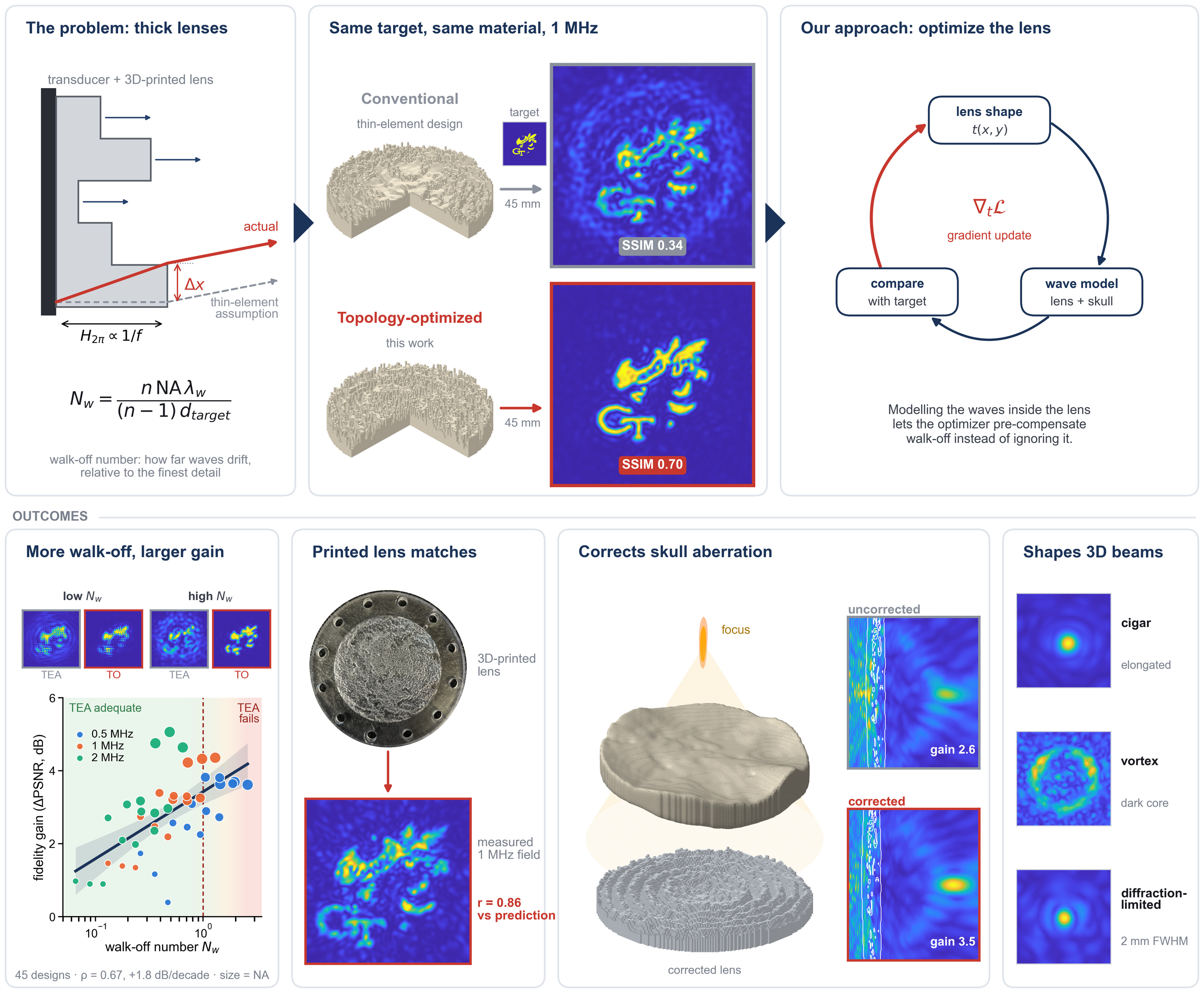}
\caption*{\textbf{Graphical abstract.} At transcranial frequencies a 3D-printed lens is millimeters thick, so sound refracts and drifts sideways inside it, and a lens designed as a thin phase screen blurs the intended pattern. Optimizing the full lens shape with a wave model that propagates through the lens and the skull sharpens the reconstruction at 1 MHz (SSIM 0.34 to 0.70), and the gain grows with the walk-off number $N_\mathrm{w}$. Printed lenses reproduce their predicted fields, correct skull aberration and shape volumetric beams.}
\end{figure*}

For the sub-megahertz to low-megahertz regime ($\lesssim$1 MHz) most relevant to transcranial applications, however, the lens thickness needed to impart a full $2\pi$ phase shift scales inversely with frequency ($L_{2\pi} \propto 1/f$) and grows substantially relative to the wavelength. Combined with the high numerical apertures needed for focused neuro-interventions, this drives the lens into an acoustically thick regime ($L \gg \lambda$), where refractive walk-off and volumetric wave propagation invalidate the thin-element approximation (TEA) and introduce significant thickness-dependent amplitude and phase errors. At higher frequencies such as 2 MHz, where $L_{2\pi}$ is comparatively small, the TEA remains a reasonable approximation, but the resulting penetration loss and dense phase wrapping limit clinical utility. The relevant figure of merit, the refractive walk-off number $N_\mathrm{w}$ introduced below, defines a frequency-dependent operating regime in which sub-MHz transcranial holograms reside, rather than a material-set thin-to-thick boundary.

Lower frequencies offer better skull penetration, while operating near 1 MHz balances the spatial resolution needed for complex holographic patterns against sufficient penetration. This range is actively being explored for high-precision, non-thermal transcranial ultrasound, including targeted neuromodulation \cite{R12,R40} and blood-brain barrier opening \cite{R13}, where spatial precision is paramount and where traditional lens design paradigms begin to fail.

High-frequency holography also requires extended optimization to prevent on- and off-target hotspots. Time-domain simulations model heterogeneities accurately but are computationally intensive, taking hours per field \cite{R8}, which makes them impractical for optimizing the large apertures required for transcranial ultrasound \cite{R14}. Frequency-domain methods are faster. The standard angular spectrum approach is fast but typically does not account for heterogeneous propagation \cite{R4}, whereas methods such as the modified mixed-domain method \cite{R15} capture strong heterogeneity and reflections yet remain difficult to embed in rapid iterative optimization. Hologram design can be accelerated with automatic differentiation \cite{R16}, and gradient-based optimization has been applied to acoustic holograms \cite{R5,R17}, but rapid design of 3D-printed lenses free of thickness-dependent amplitude and phase errors that simultaneously accounts for aberration in heterogeneous media remains an important challenge \cite{R18}.

Here we introduce a unified framework for high-fidelity acoustic holography that designs the lens topology directly. It builds on HASA, a fast spectral method for wave propagation in complex media \cite{R19}, which incorporates in-plane variations in speed of sound and supports differentiable optimization with the ADAM optimizer. The novelty of this HASA-ADAM framework \cite{R20,R21} is direct topology optimization: unlike phase-only methods that rely on the TEA, it accounts for the amplitude and phase changes that accumulate inside the lens material itself, which is critical at megahertz frequencies. Using simulations across frequency, numerical aperture and lens material, we show that modelling propagation inside the lens improves fidelity over TEA-based designs in every configuration tested, most strongly at or below 1 MHz. We then validate printed lenses with hydrophone measurements in the free field and show that patient-specific lenses correct skull-induced aberration, generalize across skull segments, and shape volumetric beams through the human skull.

\section{Results}

\subsection{A differentiable framework that optimizes lens topology}

\begin{figure*}[p]
\centering
\includegraphics[width=\textwidth]{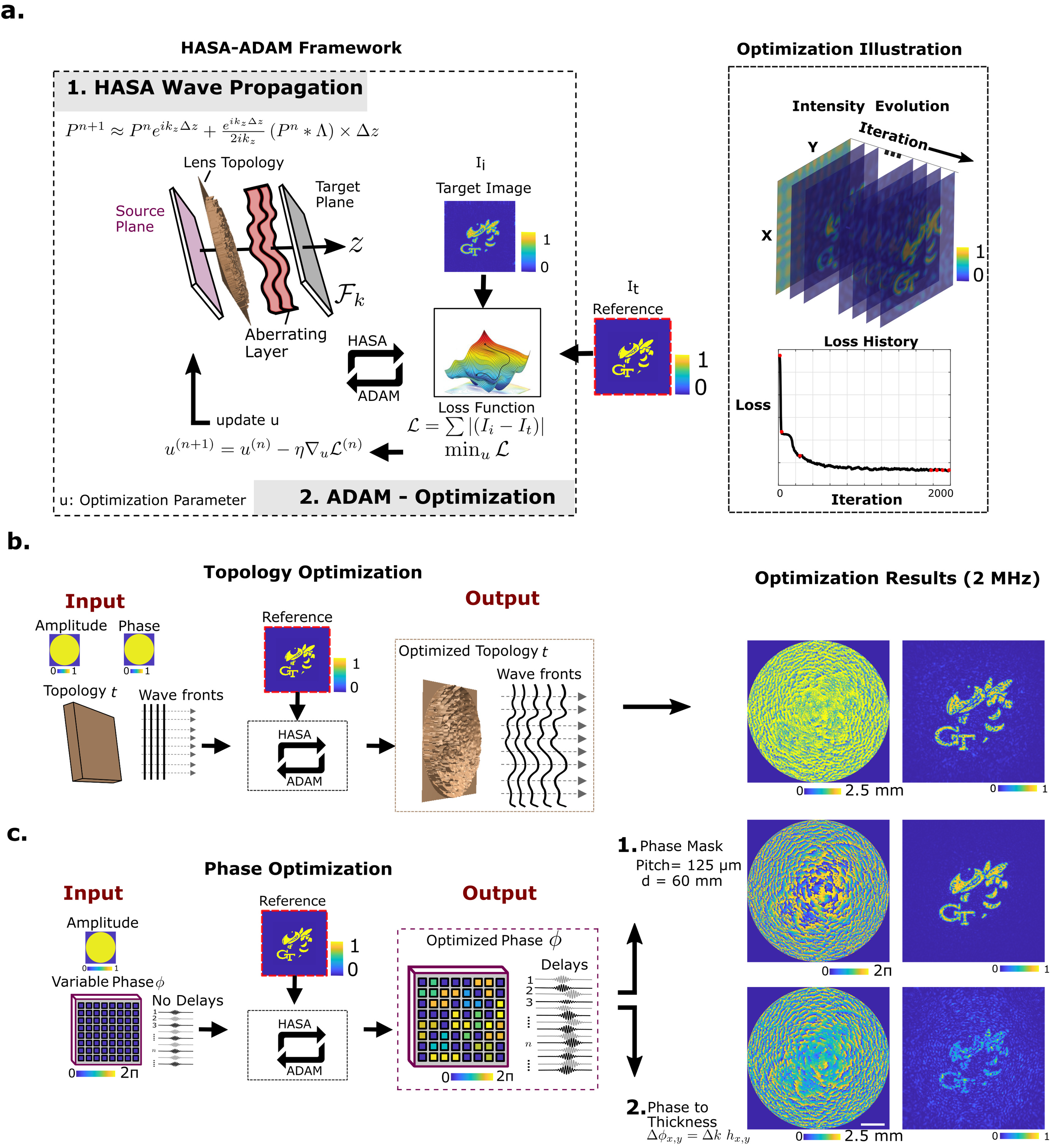}
\caption{\textbf{A unified differentiable framework that optimizes 3D lens topology rather than a 2D phase screen.} (\textit{a}) The pipeline couples a heterogeneous angular spectrum (HASA) forward solver to an ADAM optimizer (Table~\ref{tab:1}), so gradients of the target-plane intensity loss flow back through the full wave-propagation graph to either the lens thickness profile (topology mode) or the source phase (phase mode). (\textit{b}) Topology optimization for a complex 2D hologram at 2 MHz: the optimizer sculpts the physical thickness map directly and reaches high-fidelity reconstruction, sidestepping the lossy phase-to-thickness conversion. (\textit{c}) Phase optimization at 2 MHz (\SI{125}{\micro\meter} pitch, 60 mm aperture) also succeeds, because the lens is acoustically thin and the thin-element approximation holds. The contrast becomes acute at lower frequencies, where only topology optimization remains viable.}
\label{fig:1}
\end{figure*}

The framework couples HASA, a fast spectral propagator for complex media \cite{R19}, to the ADAM optimizer, which minimizes the absolute difference in intensity between a target image and the reconstructed image plane \cite{R16} (Fig.~\ref{fig:1}a and Table~\ref{tab:1}). To address the stochastic nature of ADAM and ensure reproducibility, we performed an empirical hyperparameter sweep (Fig.~\ref{fig:S1} and Table~\ref{tab:S1}). A learning rate $\eta = 0.01$ and a total-variation regularizer weight $\lambda = 0.01$ gave the best convergence and smoothed the lens geometry for manufacturability without compromising acoustic accuracy. The approach exploits the ability of HASA to carry in-plane varying speed-of-sound maps and to support differentiable optimization of the lens thickness profile, assuming an initial zero phase and constant amplitude at the source. It therefore performs direct topology optimization, designing lens topologies that account for wave propagation within the lens and encode complex acoustic holograms in the megahertz range (Fig.~\ref{fig:1}b). The same machinery can instead optimize the source phase (Fig.~\ref{fig:1}c).

\begin{table*}[t]
\centering
\caption{HASA-ADAM algorithm for (\textit{a}) topology optimization and (\textit{b}) phase optimization}
\label{tab:1}
\begin{minipage}[t]{0.47\textwidth}
\textbf{(\textit{a}) Topology optimization}
\begin{algorithmic}[1]
\Require $N_\mathrm{max}, \epsilon, \eta, \lambda, u^{(0)}, I_{t}$
\Ensure $t, \mathcal{L}$
\While{$n < N_\mathrm{max}$ and $\mathcal{L}^{(n)} > \epsilon$}
  \State $t^{(n)} \gets \min\{\mathrm{softplus}(u^{(n)}),\, t_\mathrm{max}\}$
  \State $P^{(n)} \gets \mathcal{H}(P_{0}, t^{(n)});\ I^{(n)} \gets |P^{(n)}|^{2}$
  \State $\mathcal{L}^{(n)} \gets \sum_{x,y}|I^{(n)} - I_{t}| + \lambda\mathcal{R}(t^{(n)})$
  \State $u^{(n+1)} \gets u^{(n)} - \eta\nabla_{u}\mathcal{L}^{(n)};\ n \gets n+1$
\EndWhile
\State \Return $t^{(n)}, \mathcal{L}^{(n)}$
\end{algorithmic}
\end{minipage}%
\hfill
\begin{minipage}[t]{0.47\textwidth}
\textbf{(\textit{b}) Phase optimization}
\begin{algorithmic}[1]
\Require $N_\mathrm{max}, \epsilon, \eta, \lambda, P_{0}, P_{t}$
\Ensure $\phi, \mathcal{L}$
\While{$n < N_\mathrm{max}$ and $\mathcal{L}^{(n)} > \epsilon$}
  \State $P_{i}^{(n)} \gets \mathcal{H}(P_{0}, \phi^{(n)});\ I_{i}^{(n)} \gets |P_{i}^{(n)}|^{2}$
  \State $\mathcal{L}^{(n)} \gets \sum_{x,y}|I_{i}^{(n)} - I_{t}| + \lambda\mathcal{R}(\phi^{(n)})$
  \State $\phi^{(n+1)} \gets \phi^{(n)} - \eta\nabla_{\phi}\mathcal{L}^{(n)}$
  \State $P_{0}^{(n+1)} \gets \alpha_{h}P_{0}^{(n)};\ n \gets n+1$
\EndWhile
\State \Return $\phi^{(n)}, \mathcal{L}^{(n)}$
\end{algorithmic}
\end{minipage}
\end{table*}

To test the concept we first generated a complex holographic pattern at 2 MHz, comparing direct topology optimization (Fig.~\ref{fig:1}b) with the conventional route of optimizing a phase pattern and converting it to a thickness profile under the TEA (Fig.~\ref{fig:1}c). Both used a discretization of $\lambda/6$ (\SI{125}{\micro\meter} at 2 MHz) over a 60 mm aperture. At 2 MHz both produced the target pattern with high fidelity, because the thickness required for a full $2\pi$ shift is small, the lens remains acoustically thin, and the TEA holds well enough.

Many practical applications, transcranial ultrasound among them, instead operate at 1 MHz or below to limit attenuation and penetrate the skull safely. As the frequency falls the wavelength grows, and a proportionally thicker lens is needed for the same phase modulation. In this regime the lens becomes acoustically thick and the TEA fails through lateral energy migration, refractive walk-off and internal volumetric propagation, so direct topology optimization becomes necessary to preserve fidelity.

The computational complexity of the framework, $\mathcal{O}(N^{2}\log N)$, is higher than that of the homogeneous angular spectrum approach, $\mathcal{O}(N\log N)$ (Fig.~\ref{fig:S2}) \cite{R4}, yet an optimized topology converges in about 20 min at $\lambda/10$ discretization at 1 MHz within a 40 mm domain on a 24 GB NVIDIA RTX 3090 GPU. Increasing the linear dimension by 50\% to 60 mm, the limit of the available GPU memory, raises the time about threefold, close to the 2.4-fold expected from the $\mathcal{O}(N^{2}\log N)$ scaling and consistent with memory overhead, which shows the method scales to clinical-scale apertures. Downsampling to $\lambda/6$ at 1 MHz cuts the time to about 15 min for 500 iterations with no loss in performance, so we adopted $\lambda/6$ throughout. HASA-ADAM therefore designs 3D-printed lenses for high-fidelity holography at clinically relevant apertures within minutes, at a fraction of the cost of a comparable phased array.

\subsection{Refractive walk-off defines the operating regime}

Conventional design relies on the TEA, under which phase shifts are imparted purely locally and lateral energy transport within the lens is neglected. As acoustic holography scales toward megahertz operation at high numerical aperture, propagation through the lens body moves from a locally planar regime to one in which the lens behaves as a volumetric refractive element. The dimensionless quantities that matter are not those of the lens material alone but of the lens together with the field it must reproduce.

\begin{figure*}[p]
\centering
\includegraphics[width=0.795\textwidth]{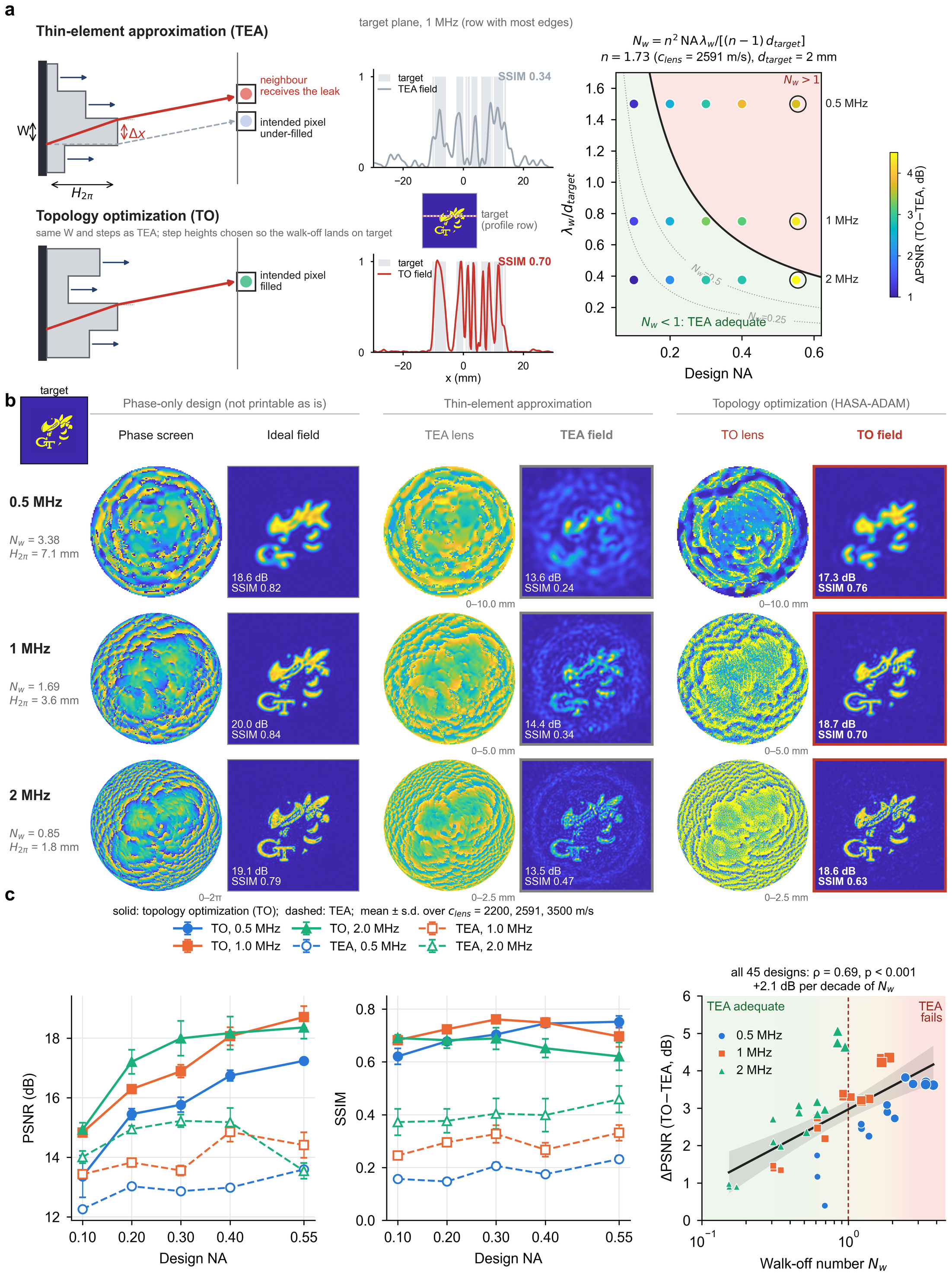}
\caption{\textbf{Accounting for refractive walk-off inside the lens preserves holographic fidelity, with the largest structural gains at transcranial frequencies at or below 1 MHz.} All panels come from one simulation sweep of the GT Bee hologram (60 mm aperture, 45 mm projection; 0.5, 1.0 and 2.0 MHz; NA 0.1 to 0.55; $c_\mathrm{lens} = 2200$, 2591 and 3500 m/s). (\textit{a}) \textit{Left}: both designs use the same pixel width $W$ and number of steps. A ray through the full-wrap step $H_{2\pi}$ travels obliquely inside the lens and refracts toward the axis at the lens-water face. Under the thin-element approximation (\textit{top}) it leaves displaced by $\Delta x$ from where that approximation places it (dashed grey), underfilling the intended pixel and leaking into its neighbour; topology optimization (\textit{bottom}) chooses step heights so that the walk-off lands on the intended pixel. \textit{Middle}: the 1 MHz field along the target row with the most edges ($\mathrm{NA} = 0.55$, $c_\mathrm{lens} = 2591$ m/s), thin-element (grey) and topology-optimized (red) over the target (shaded); inset, the target with this row marked. \textit{Right}: regime diagram in $(\mathrm{NA}, \lambda_\mathrm{w}/d_\mathrm{target})$ space for $n = 1.73$ and $d_\mathrm{target} = 2$ mm; black curve, $N_\mathrm{w} = 1$; dotted curves, $N_\mathrm{w} = 0.25$ and 0.5. Markers are the 15 swept designs at this $c_\mathrm{lens}$, coloured by PSNR gain; circled markers appear in \textit{b}. (\textit{b}) $\mathrm{NA} = 0.55$ designs at 0.5, 1.0 and 2.0 MHz: optimized phase screen and its ideal field, the thin-element lens from phase-to-thickness conversion and its field, and the topology-optimized lens and its field, with PSNR and SSIM relative to the target (inset). Thickness and phase share one colour scale per row. (\textit{c}) PSNR (\textit{left}) and SSIM (\textit{middle}) versus design NA for topology optimization (solid, filled) and the thin-element approximation (dashed, open), mean $\pm$ s.d. over the three lens sound speeds (per-speed data, Fig.~\ref{fig:S8}; full reconstructions, Figs.~\ref{fig:S5} to \ref{fig:S7}). \textit{Right}: PSNR gain versus walk-off number for all 45 designs (colour, frequency; marker size, NA), with a least-squares fit in $\log N_\mathrm{w}$ and its 95\% band (Spearman $\rho = 0.69$, $P < 0.001$). Topology optimization exceeds the thin-element approximation in all 45 combinations (mean $\Delta$PSNR 2.8 dB; mean $\Delta$SSIM 0.52, 0.43 and 0.26 at 0.5, 1.0 and 2.0 MHz). Movies S1 and S2.}
\label{fig:2}
\end{figure*}

The geometric origin of the failure is shown in Fig.~\ref{fig:2}a. A refractive hologram of relative index $n = c_\mathrm{lens}/c_\mathrm{w}$ at frequency $f$ in water has two intrinsic length scales: the full-wrap height $H_{2\pi} = \lambda_\mathrm{w}/(1 - c_\mathrm{w}/c_\mathrm{lens}) = n\lambda_\mathrm{w}/(n-1)$ required for a $2\pi$ phase shift, and the lateral feature width $W = \lambda_\mathrm{w}/(2\,\mathrm{NA})$ set by the design numerical aperture, with $\lambda_\mathrm{w} = c_\mathrm{w}/f$. Both scale as $1/f$, so the geometric aspect ratio

\begin{equation}
\mathrm{AR} = \frac{H_{2\pi}}{W} = \frac{2\,n\,\mathrm{NA}}{n-1}
\label{eq:1}
\end{equation}

\noindent is frequency-independent and fixed once the lens material and design NA are chosen ($\mathrm{AR} \approx 2.6$ for our parameters, $\mathrm{NA} = 0.55$ and $n = 1.73$). The same holds for the dimensionless refractive walk-off $\Delta x / W = 2\,n\,\mathrm{NA}\tan\theta_\mathrm{r}/(n-1)$, where $\theta_\mathrm{r} = \arcsin(n\,\mathrm{NA})$ is the propagation angle inside the lens. This follows from Snell's law at the lens-water face: the transverse wavenumber is conserved, and because the resin is faster than water ($n > 1$) the angle inside the lens exceeds the exit angle, real for $n\,\mathrm{NA} < 1$. No two-scale geometric criterion produces a frequency-dependent thin-to-thick boundary; defining the thick regime by $H \gg \lambda$ gives a material-set answer that is the same at every operating frequency.

A frequency-dependent regime requires a third length that does not scale with $\lambda_\mathrm{w}$. In holographic therapy that length is the smallest target-field feature $d_\mathrm{target}$, whether a focal-spot separation, a lesion diameter or an anatomical detail, set by the application and fixed in absolute units as $f$ varies. The controlling dimensionless number is the refractive walk-off number

\begin{equation}
N_{\mathrm{w}} = \frac{\Delta x}{d_{\mathrm{target}}}
= \frac{n^{2} \cdot \mathrm{NA} \cdot \lambda_{\mathrm{w}}}{(n-1)\, d_{\mathrm{target}}},
\label{eq:2}
\end{equation}

\noindent which crosses unity at a crossover frequency $f_\mathrm{c} = n^{2}\,\mathrm{NA}\,c_\mathrm{w}/[(n-1)d_\mathrm{target}]$. Here $\Delta x$ is taken to first order in NA: at the lens-water face the transverse wavenumber is conserved, so a component that leaves at $\sin\theta_\mathrm{w} = \mathrm{NA}$ travels inside the faster resin at $\sin\theta_\mathrm{r} = n\,\mathrm{NA}$, and $\Delta x = H_{2\pi}\tan\theta_\mathrm{r} \approx n^{2}\,\mathrm{NA}\,\lambda_\mathrm{w}/(n-1)$. For our geometry ($\mathrm{NA} = 0.55$, $n = 1.73$) and $d_\mathrm{target} = 2$ mm, $f_\mathrm{c} \approx 1.7$ MHz; for the roughly 1 mm finest strokes of the GT Bee target, $f_\mathrm{c} \approx 3.4$ MHz. Above $f_\mathrm{c}$ the walk-off is small relative to the target and the TEA yields acceptable fidelity; below it the walk-off becomes comparable to or exceeds $d_\mathrm{target}$ and the TEA fails. The boundary is operational: it depends on what the lens is asked to reproduce, not on the material. Fig.~\ref{fig:2}a (right) shows the resulting regime diagram in $(\mathrm{NA}, \lambda_\mathrm{w}/d_\mathrm{target})$ space, with the $N_\mathrm{w} = 1$ contour separating the TEA-sufficient regime from the operating regime in which sub-MHz transcranial holograms reside, together with the 15 swept designs for $c_\mathrm{lens} = 2591$ m/s coloured by their PSNR gain. The first-order form keeps $N_\mathrm{w}$ finite for every design; the full expression diverges as $n\,\mathrm{NA} \rightarrow 1$, a limit exceeded only by the three designs with $c_\mathrm{lens} = 3500$ m/s and $\mathrm{NA} = 0.55$.

The mechanism itself is explicit in Fig.~\ref{fig:2}a: physical waves refract according to Snell's law as they traverse the steep topographical gradients of the sawtooth lens features, with internal angle $\theta_\mathrm{r} = \arcsin(n\,\mathrm{NA})$. Instead of propagating longitudinally as the TEA assumes, energy migrates laterally within the lens by $\Delta x = H_{2\pi}\tan\theta_\mathrm{r}$ and exits displaced from the design coordinate. Energy intended for one coordinate is diverted into adjacent pixels, producing the deterministic loss of field fidelity that 2D phase-projection models neglect. Walk-off is not specific to TEA-designed lenses: a topology-optimized lens with the same pixel width $W$ refracts energy in the same way, but because the forward model propagates the field through the lens volume, the optimizer selects a topology whose walk-off delivers each pixel's energy to its intended coordinate (Fig.~\ref{fig:2}a, bottom).

\subsection{Topology optimization gains most where walk-off is largest}

To assess this mechanism and the resulting frequency dependence, we compared TEA-based phase optimization with HASA-ADAM topology optimization across operating frequencies (Fig.~\ref{fig:2}b). Converting an idealized 2D phase map into a physical 3D geometry degrades the reconstruction at every frequency, and increasingly so as $N_\mathrm{w}$ grows. Topology optimization instead optimizes the continuous physical thickness profile with a full-wave heterogeneous propagator that deterministically pre-compensates for walk-off and the associated internal propagation.

At 0.5 MHz the full-wrap height is 7.1 mm and walk-off is several times the target scale ($N_\mathrm{w} \approx 3.4$ for $d_\mathrm{target} = 2$ mm). The TEA lens degrades sharply (PSNR 13.6 dB, SSIM 0.24) while topology optimization recovers most of the structure (17.3 dB, SSIM 0.76), close to the ideal phase screen (18.6 dB, SSIM 0.82). At 1.0 MHz ($N_\mathrm{w} \approx 1.7$, inside the operating regime) the TEA lens reaches 14.4 dB (SSIM 0.34) and topology optimization 18.7 dB (SSIM 0.70). At 2.0 MHz, just below the regime boundary ($N_\mathrm{w} \approx 0.85$), dense phase wrapping that resets every $\lambda_\mathrm{w} \approx 0.75$ mm introduces edge diffraction and localized scattering at the wrap discontinuities; even here topology optimization (18.6 dB, SSIM 0.63) outperforms the TEA lens (13.5 dB, SSIM 0.47).

Two features of this high-NA dataset stand out. First, the structural similarity of the topology-optimized hologram to the target improves as the operating frequency decreases (SSIM 0.63, 0.70 and 0.76 at 2.0, 1.0 and 0.5 MHz), whereas that of the TEA lens falls (0.47, 0.34 and 0.24), so the structural gain from modelling the lens grows from 0.16 to 0.52 as walk-off increases. Second, PSNR does not improve with frequency: the 1.0 MHz topology-optimized hologram (18.7 dB) matches the 2.0 MHz one (18.6 dB) at half the frequency and twice the wavelength. Across the NA sweep (Fig.~\ref{fig:2}c) the 2.0 MHz design leads in PSNR for $\mathrm{NA} \leq 0.3$, the 1.0 MHz design matches it at $\mathrm{NA} = 0.4$ and exceeds it at $\mathrm{NA} = 0.55$. Together these results identify the sub-megahertz to low-megahertz range as a favourable operating window for high-NA, high-fidelity acoustic holography, provided walk-off is modelled during design.

Full reconstruction sweeps across all sampled (NA, $c_\mathrm{lens}$) combinations at each frequency are given in Figs.~\ref{fig:S5} to \ref{fig:S7}. Fig.~\ref{fig:2}c condenses them by averaging over the three lens sound speeds, whose effect is minor (s.d.~$\leq$ 0.6 dB in PSNR; Fig.~\ref{fig:S8}). Topology optimization exceeds the TEA on both PSNR and SSIM in all 45 combinations sampled. Its PSNR rises with NA at every frequency, whereas the TEA gains at most 1.4 dB across the NA range and declines beyond $\mathrm{NA} = 0.4$ at 1.0 and 2.0 MHz. The PSNR gain is similar across frequencies (mean $\Delta$PSNR 2.8, 2.9 and 2.8 dB at 0.5, 1.0 and 2.0 MHz), but the structural gain grows as the frequency decreases (mean $\Delta$SSIM 0.52, 0.43 and 0.26). Across all 45 designs the PSNR gain increases with the walk-off number (Spearman $\rho = 0.69$, $P < 0.001$; 2.1 dB per decade of $N_\mathrm{w}$; Fig.~\ref{fig:2}c, right). The trend is not a frequency effect: it holds within each frequency, where $N_\mathrm{w}$ changes through the design NA and the lens sound speed ($\rho = 0.84$, 0.86 and 0.87 at 0.5, 1.0 and 2.0 MHz; $P < 0.001$). The structural gain, by contrast, is set mainly by frequency and does not increase with $N_\mathrm{w}$ within a frequency. Accounting for propagation inside the lens therefore matters most at the frequencies at or below 1 MHz used for transcranial ultrasound.

Parametric sweeps corroborate these trends. Movie S1 sweeps the frequency from 5.0 to 0.5 MHz, showing the lens height growing as $1/f$ and walk-off growing with it in absolute units against a fixed target. Movie S2 isolates focusing strength, showing that higher NA demands steeper internal refraction angles and therefore reaches deeper into the operating regime at fixed $f$. We also compared HASA-ADAM against the conventional iterative angular spectrum approach (IASA) on benchmark holograms (Fig.~\ref{fig:S4}). At 1 MHz and below, standard IASA phase optimization fails entirely, with incoherent scattering and substantial PSNR loss, whereas topology optimization pre-compensates for walk-off and yields coherent, high-fidelity reconstructions.

\subsection{Printed lenses reproduce their designed fields}

Having established the breakdown of the TEA, we validated the framework experimentally. We fabricated two 1 MHz holographic lenses by 3D printing and characterized their acoustic fields at the design plane with a hydrophone raster scan. The measurements were compared with the field predicted for the printed lens geometry by the HASA forward model, which agreed with a k-Wave simulation of the same lens type at $r = 0.84$ (\emph{Materials and Methods}). To isolate the dependence of fidelity on hologram feature size, we evaluated two GT Bee holograms encoding the same target at different physical scales: a large hologram whose smallest features remain well above the acoustic wavelength, and a small hologram whose features approach the wavelength and therefore sit deeper into the operating regime defined by $N_\mathrm{w}$.

\begin{figure*}[t]
\centering
\includegraphics[width=\textwidth]{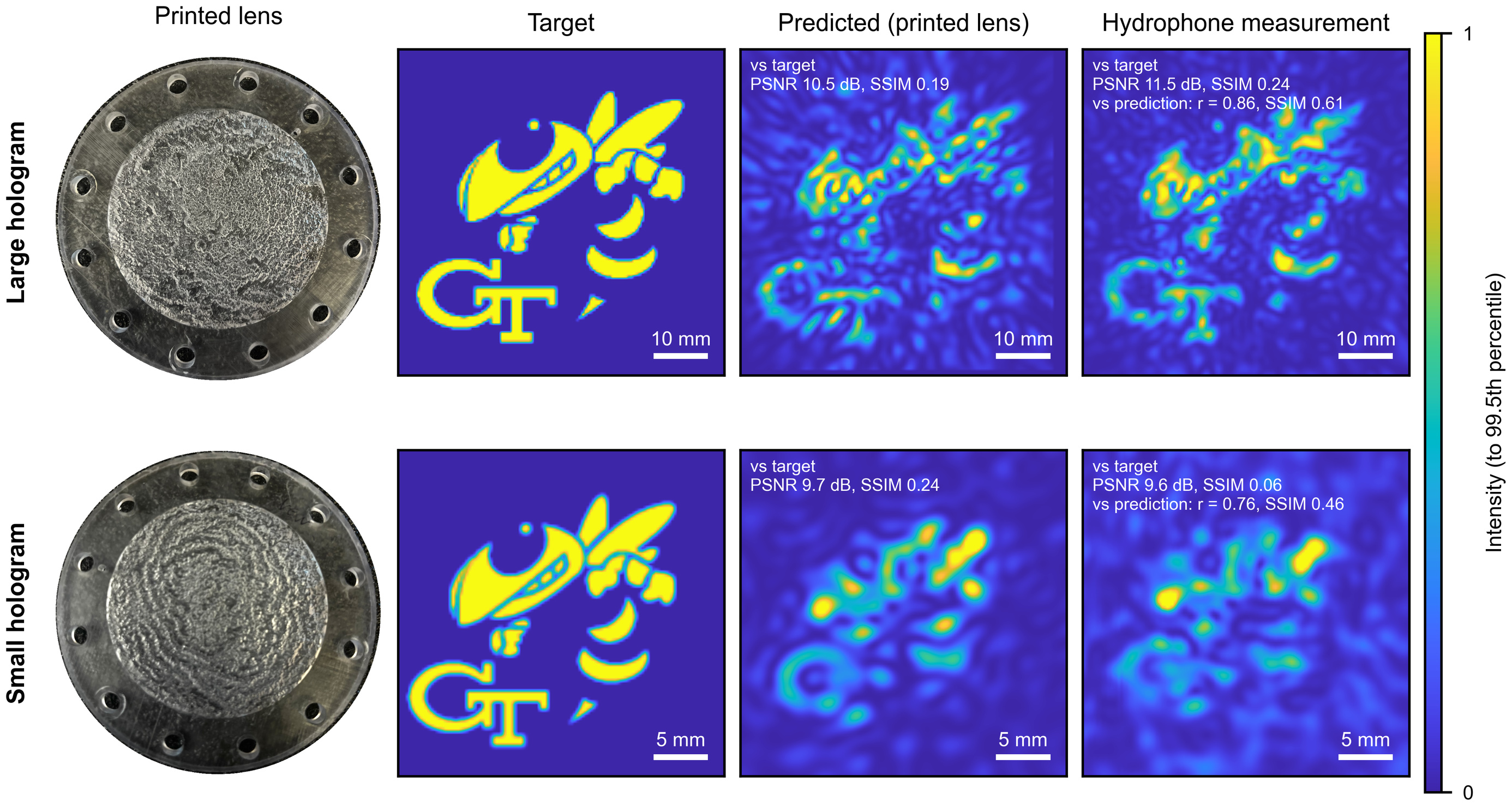}
\caption{\textbf{Free-field experimental validation of topology-optimized lenses at 1 MHz.} \textit{Rows}: large GT Bee hologram, whose smallest features remain well above the wavelength (\textit{top}), and small hologram, whose features approach the wavelength (\textit{bottom}). \textit{Columns}: photograph of the printed lens, design target, field predicted for the printed lens geometry at the 45 mm design plane by the HASA forward model, and hydrophone measurement with a 0.2 mm needle hydrophone (large, 60 $\times$ 60 mm scan at 0.5 mm step; small, 30 $\times$ 30 mm scan at 0.25 mm step). Metrics use intensity normalized to the peak of each map; for display, each field is scaled to its 99.5th percentile on the same colour scale, so that isolated speckle hot spots do not set the colour range. The prediction and target were registered to the measurement by in-plane translation only. Insets give PSNR and SSIM relative to the target and, for the measurement, the Pearson correlation $r$ and SSIM relative to the prediction. The large hologram's measurement closely follows its prediction ($r = 0.86$), whereas the small hologram's agrees less well ($r = 0.76$) and falls below its prediction in fidelity to the target. Scale bars: 10 mm (\textit{top}), 5 mm (\textit{bottom}).}
\label{fig:3}
\end{figure*}

Both lenses reproduced the overall GT Bee morphology (Fig.~\ref{fig:3}). For the large hologram the measured field closely followed the prediction for the printed lens (Pearson $r = 0.86$, SSIM 0.61), and its fidelity to the target (PSNR 11.5 dB, SSIM 0.24, $r = 0.83$) matched or exceeded the prediction (PSNR 10.5 dB, SSIM 0.19, $r = 0.77$). For the small hologram, agreement with the prediction was lower ($r = 0.76$, SSIM 0.46) and the measured fidelity to the target (PSNR 9.6 dB, SSIM 0.06, $r = 0.43$) fell below the prediction (PSNR 9.7 dB, SSIM 0.24, $r = 0.50$), with reduced edge definition and a diffuse background haze that the compressional model does not reproduce. In both cases the measured field is smoother than the prediction: convolving the prediction with a Gaussian of $\sigma = 0.5$ mm raises the agreement to $r = 0.89$ (large) and 0.78 (small) and, for the large hologram, matches the fraction of the field above half maximum (3.9\% versus 3.7\%). This indicates an effective measurement resolution of about 1.2 mm FWHM, coarser than the 0.2 mm hydrophone, which may arise from positioning error, the finite bandwidth of the excitation burst and printing tolerances. Absolute PSNR and SSIM against the binary target are low for both prediction and measurement, because the acoustic field forms speckle-modulated edges rather than uniformly filled regions; the metrics are most informative compared between conditions. The growing gap between prediction and measurement as the smallest target feature approaches the wavelength is consistent with the walk-off argument above, although other unmodelled effects may also contribute.

These measurements show that a 3D-printed lens designed with HASA-ADAM reproduces the intended pattern in practice when the target features remain well above the wavelength. Because both lenses were printed and measured with the same process, they also suggest that hologram feature size relative to the wavelength, rather than fabrication quality, limits the achievable fidelity, although the two lenses come from different design iterations and a matched pair would make this comparison conclusive. The physical origin of the residual discrepancy at small feature sizes, including a possible contribution from elastic mode conversion at the lens-water interface, is considered in the \emph{Discussion} and in Fig.~\ref{fig:S11}.

\subsection{Topology optimization improves aberration correction through the human skull}

\begin{figure*}[p]
\centering
\includegraphics[width=0.84\textwidth]{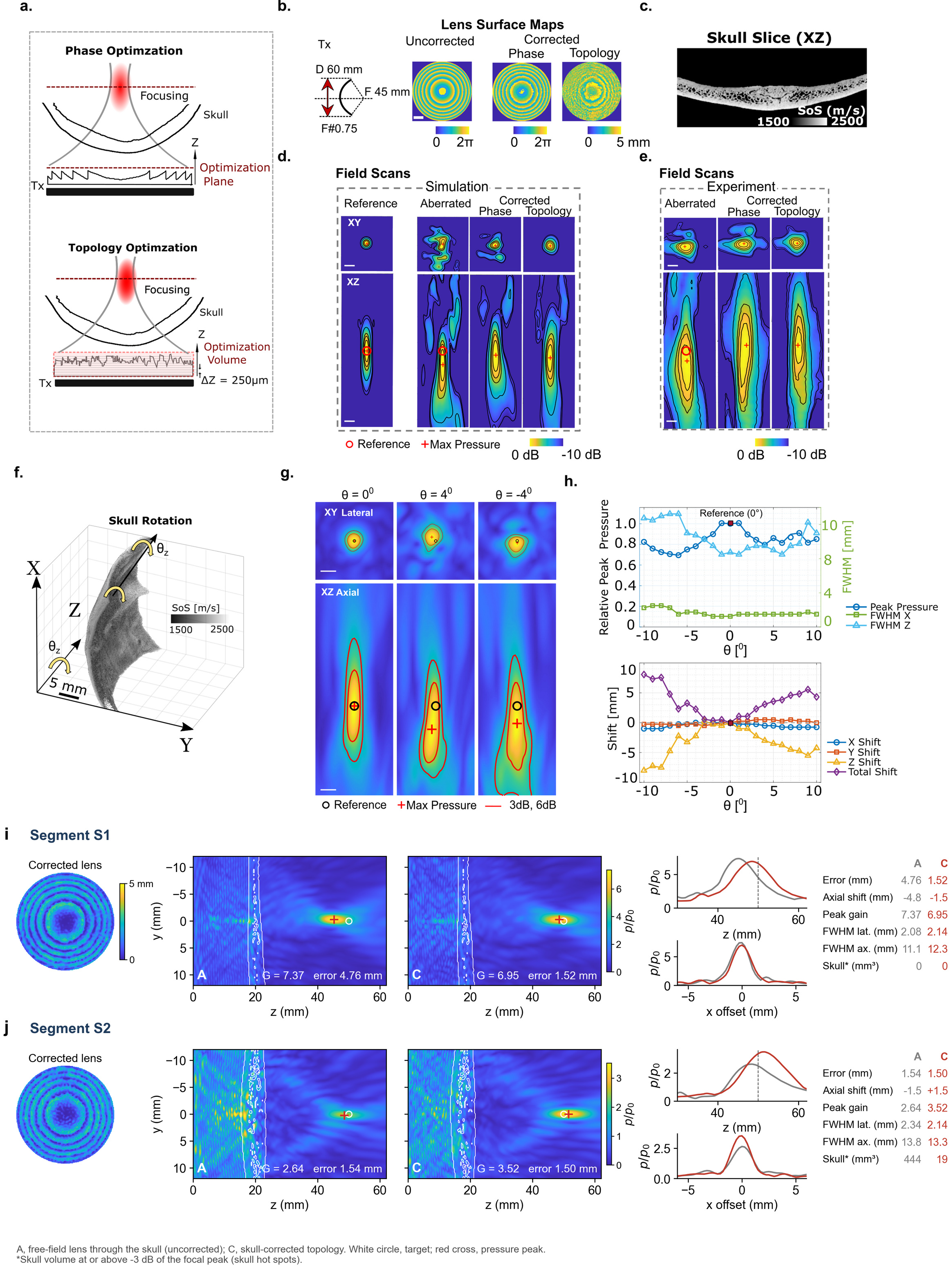}
\caption{\textbf{Through a human skull, topology optimization outperforms aberration-corrected phase, which outperforms the uncorrected baseline, and the correction generalizes across skull segments.} (\textit{a}) Phase versus topology optimization for single-focus trans-skull targeting. (\textit{b}) Phase maps for a 60 mm, $F\#0.75$ transducer with and without skull correction, plus the equivalent topology map (scale bar 1 cm). (\textit{c}) Skull speed-of-sound map from micro-CT. (\textit{d}) Simulated and (\textit{e}) measured focal profiles for both lens types, with and without aberration correction (scale bar 2 mm); the measured axial ($xz$) scans of the two corrected lenses are rotated by 180$^\circ$ to match the simulations in \textit{d}. The corrected-topology lens recovers on-target focusing, reduces the lateral 3 dB beamwidth by 24.5\% and lowers the peak sidelobe relative to corrected phase. (\textit{f}) Skull-rotation simulation mask (scale bar 5 mm). (\textit{g}) Targeting and focusing degradation maps for $\pm$4$^\circ$ rotation (scale bar 2 mm). (\textit{h}) Effect of rotation on peak amplitude and FWHM (\textit{top}) and on focal-spot shift in $x$, $y$ and $z$ (\textit{bottom}); sub-degree alignment errors translate into millimeter-scale focal shifts, which makes skull-lens registration the practical bottleneck for clinical translation. (\textit{i}, \textit{j}) Generalization to skull segments S1 (\textit{i}) and S2 (\textit{j}) in 3D k-Wave simulations: corrected lens thickness; coronal ($y$-$z$) pressure of the free-field lens on the skull (A) and of the skull-corrected topology (C) on a common scale, with the skull outlined, the target marked by a white circle and the pressure peak by a red cross; axial and lateral profiles through the peak; and focal metrics. In S1, correction reduces the targeting error from 4.76 to 1.52 mm; in S2, it raises the peak gain from 2.64 to 3.52 and reduces the skull volume at or above $-$3 dB of the focal peak from 444 to 19 mm$^{3}$. Metrics for all three segments are in Table~\ref{tab:2}.}
\label{fig:4}
\end{figure*}

Moving from idealized free-field holograms to patient-specific use, we designed optimized topologies for a single focus at 1 MHz with F-number 0.75 through the human skull (Fig.~\ref{fig:4}a and b). Using micro-CT derived maps of spatially varying speed of sound and density (Fig.~\ref{fig:4}c), the framework intrinsically pre-compensates for skull-induced wavefront aberration. We compared the thickness-optimized lens (corrected topology) with a lens designed for a single focus in the free field (uncorrected) and with a lens optimized for aberration but realized through standard phase-to-thickness conversion (corrected phase). Pressures were compared in acoustic simulation and in ex vivo trans-skull experiments (Fig.~\ref{fig:4}d). The framework not only corrected the aberration but also substantially reduced sidelobes in both the axial ($xz$) and lateral ($xy$) focal planes, in both measurement and simulation.

Quantitatively, corrected topology achieved a marked reduction in peak sidelobe level relative to corrected phase, showing the advantage of direct topology optimization in minimizing off-target exposure. It also achieved a 24.5\% reduction in lateral 3 dB beam width and realigned the focal spot with the reference, in contrast to the shift seen without aberration correction (Fig.~\ref{fig:4}d). The focal fields of the other two lenses showed stronger aberration and higher sidelobes.

These data show that the framework corrects skull-induced aberration and approaches diffraction-limited performance; they also indicate that the focus attained with the experimental system falls below theoretical limits (Fig.~\ref{fig:4}e). The discrepancy between simulation and experiment is most likely due to registration error, uncertainty in the skull and lens material properties, or both. Our past work has shown that skull properties must deviate by more than 20\% to cause significant errors \cite{R22}, which is unrealistic in most cases.

To investigate this, we performed a sensitivity analysis on the speed of sound (Table~\ref{tab:S2}). A $\pm$15\% mismatch in the skull or lens sound speed can produce a $\pm$30\% variation in peak focal pressure and an axial focal shift of up to 2.0 mm, consistent with the broadening observed experimentally (Fig.~\ref{fig:4}e). Mechanical rotation of the skull fixture also induces a coupled lateral translation, about 10.6 mm for a 6$^\circ$ rotation, which likely smeared the focal spot during the scan.

Residual errors may still arise from registration, so we turned to skull-lens alignment. We ran additional simulations rotating the skull about the $z$ axis in the transverse plane in increments of $\pm$1$^\circ$ (Fig.~\ref{fig:4}f) and compared the baseline axial and lateral fields (Fig.~\ref{fig:4}g). The 3 dB FWHM area grows in both the lateral and axial directions (Fig.~\ref{fig:4}g and h, top), consistent with the degradation in focusing performance in Fig.~\ref{fig:4}e, and focusing quality degrades with rotation along with a significant shift in the axial direction (Fig.~\ref{fig:4}g and h, bottom). Together these findings show that the optimized design corrects skull-induced aberration effectively and that combining it with accurate skull-lens alignment is critical for optimal focus with systems at or above 0.7 MHz.

\subsection{Correction generalizes across skull anatomies}

A single focusing trial is insufficient for clinical validation, because human skulls vary widely in thickness and curvature. We therefore designed corrected lenses for three further skull segments (S1 to S3) and evaluated them in 3D k-Wave simulations against the free-field lens placed on the same segment (Fig.~\ref{fig:4}i and j, and Table~\ref{tab:2}). Without correction, each segment produced a different failure mode. Segments S1 and S3 displaced the focus by 4.76 and 3.82 mm, almost entirely along the beam axis, while segment S2, the most aberrating, weakened the focus (peak gain 2.64, against 7.37 and 8.21 for S1 and S3), so that 444 mm$^{3}$ of skull bone reached at least $-$3 dB of the focal pressure.

\begin{table*}[t]
\centering
\caption{Focal performance through three skull segments without (A) and with (C) topology correction}
\label{tab:2}
\begin{tabular}{@{}lcccccc@{}}
\toprule
& \multicolumn{2}{c}{S1 (Fig.~\ref{fig:4}\textit{i})} & \multicolumn{2}{c}{S2 (Fig.~\ref{fig:4}\textit{j})} & \multicolumn{2}{c}{S3} \\
\cmidrule(lr){2-3}\cmidrule(lr){4-5}\cmidrule(lr){6-7}
Metric & A & C & A & C & A & C \\
\midrule
Peak gain ($p/p_{0}$)                & 7.37  & 6.95  & 2.64  & 3.52  & 8.21  & 8.52  \\
Targeting error (mm)                 & 4.76  & 1.52  & 1.54  & 1.50  & 3.82  & 2.81  \\
Axial shift (mm)                     & $-$4.75 & $-$1.50 & $-$1.50 & $+$1.50 & $-$3.75 & $-$2.75 \\
Lateral FWHM (mm)                    & 2.08  & 2.14  & 2.34  & 2.14  & 1.99  & 1.99  \\
Axial FWHM (mm)                      & 11.08 & 12.29 & 13.85 & 13.34 & 11.31 & 11.29 \\
Skull volume $\geq -3$ dB (mm$^{3}$) & 0     & 0     & 444   & 19    & 0     & 0     \\
\bottomrule
\end{tabular}

\addtabletext{A, free-field lens on the skull (uncorrected); C, skull-corrected topology. Peak gain is the focal peak pressure normalized to the source pressure, and targeting error is the distance from the peak to the target. Lateral and elevational shifts were at or below 0.5 mm in every case, so only the axial shift is listed. Skull volume is the volume of bone at or above $-$3 dB of the focal peak. The $-$3 dB focal main lobe was 8.6 to 11.1 mm$^{3}$ in all cases. 3D k-Wave simulations at 1 MHz.}
\end{table*}

Topology optimization reduced each of these errors. For S1 it cut the targeting error from 4.76 to 1.52 mm by correcting most of the axial shift, from $-$4.75 to $-$1.50 mm, with a 6\% lower peak gain. For S2 it raised the peak gain by 33\%, from 2.64 to 3.52, and reduced the skull volume at or above $-$3 dB of the focal pressure from 444 to 19 mm$^{3}$, while the focal main lobe kept a similar size (11.1 and 10.5 mm$^{3}$). For S3 it raised the peak gain from 8.21 to 8.52 and reduced the targeting error from 3.82 to 2.81 mm, although a 2.75 mm axial offset remained. Lateral and elevational shifts stayed at or below 0.5 mm in every case, and the lateral FWHM stayed between 1.99 and 2.34 mm. The residual axial offsets in S1 and S3 are comparable to the axial shifts produced by a 15\% speed-of-sound mismatch (Table~\ref{tab:S2}), and part of them may also arise from the validation geometry (\emph{Materials and Methods}).

\subsection{Multiplane optimization shapes volumetric beams}

In conventional acoustic holography the inverse problem is formulated on a single 2D plane, maximizing pressure or matching a target intensity at a fixed depth. This planar approach is cheap but assumes that optimizing the field at the target implicitly optimizes the beam path, an assumption that frequently fails in media as heterogeneous as the human skull. A loss defined only at the target slice leaves the optimizer blind to the field outside that plane, with no incentive to suppress energy accumulation before or after the focus. The result is often a needle-like focus with poor axial confinement that spreads energy over a large volume, reduces peak intensity at the target and increases the risk of pre-focal heating.

We therefore implemented a fully differentiable multiplane scheme. Volumetric gradient descent computes propagation through the entire region of interest (\emph{Materials and Methods}), which also allows arbitrary 3D pressure distributions matched to anatomical constraints. Deep-brain targets such as the hypothalamus, the dentate-rubro-thalamic tract or the anterior limb of the internal capsule are elongated structures 10 to 20 mm long rather than spherical points, and a diffraction-limited focus covers only a fraction of such a target, requiring mechanical raster scanning that adds procedure time and complexity. We therefore solved the inverse source problem for an anisotropic 3D Gaussian target.

\begin{figure*}[t]
\centering
\includegraphics[width=\textwidth]{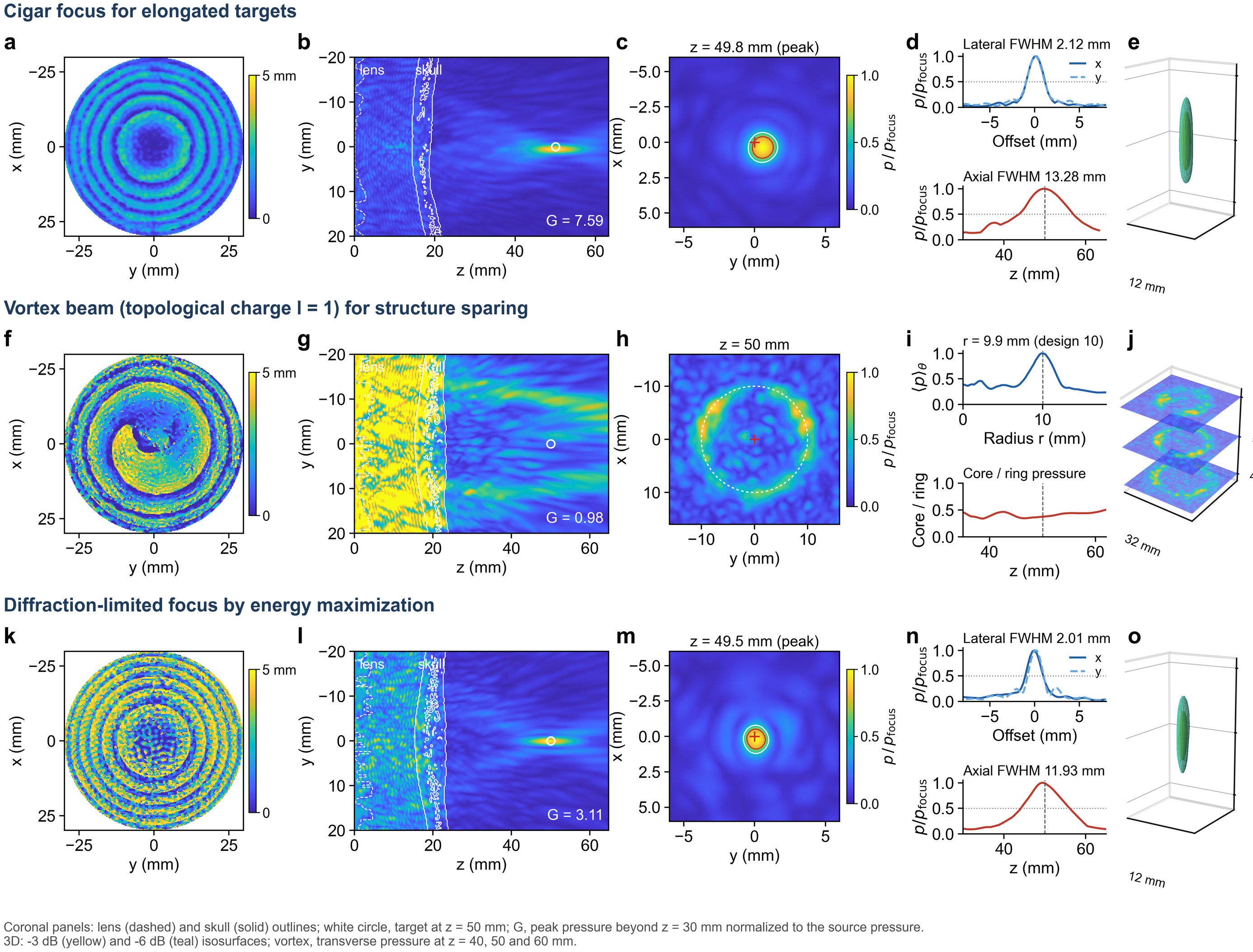}
\caption{\textbf{Volumetric optimization shapes beams through the human skull.} 3D k-Wave simulations at 1 MHz (60 mm aperture, target at $z = 50$ mm from the source plane), one row per beam type. \textit{Columns}: lens thickness; coronal ($y$-$z$) pressure through the target axis with the lens (dashed) and skull (solid) outlined and the target marked by a white circle; transverse ($x$-$y$) pressure; profiles; and 3D isosurfaces ($-$3 dB, yellow; $-$6 dB, teal). All maps use the parula colour scale. Pressure is normalized to the focal peak beyond $z = 30$ mm ($p_\mathrm{focus}$), $G$ is that peak normalized to the source pressure, and values in the lens and skull above $p_\mathrm{focus}$ are clipped. (\textit{a} to \textit{e}) Cigar focus for elongated targets, designed through skull segment S4: (\textit{c}) transverse pressure at the peak with $-$3 dB (red) and $-$6 dB (white) contours; (\textit{d}) lateral ($x$, $y$) and axial profiles through the peak (lateral FWHM 2.12 mm, axial FWHM 13.28 mm, $G = 7.59$, targeting error 0.61 mm); (\textit{e}) $-$3 dB and $-$6 dB isosurfaces. (\textit{f} to \textit{j}) Vortex beam (topological charge $l = 1$, design ring radius 10 mm) through segment S2, initialized with a spiral Fresnel warm start: (\textit{h}) transverse pressure at $z = 50$ mm with the design radius dashed; (\textit{i}) azimuthally averaged radial profile at $z = 50$ mm (\textit{top}; ring radius 9.9 mm) and core-to-ring pressure ratio along $z$ (\textit{bottom}); (\textit{j}) transverse pressure at $z = 40$, 50 and 60 mm, stacked to show the ring and its darker core along the beam. The ring is non-uniform and weak relative to the source ($G = 0.98$). (\textit{k} to \textit{o}) Diffraction-limited focus through segment S2, obtained by maximizing the intensity inside a sub-wavelength mask rather than matching a Gaussian target (lateral FWHM 2.01 mm, about $1.34\lambda$; axial FWHM 11.93 mm; $G = 3.11$; targeting error 0.56 mm). Metrics are listed in Table~\ref{tab:3}.}
\label{fig:5}
\end{figure*}

Enforcing this volumetric constraint penalizes energy outside the designated ellipsoid. A cigar-focus lens designed through a fourth skull segment (S4; Fig.~\ref{fig:5}a) produced, in k-Wave validation, an elongated focus 0.61 mm from the target with a lateral FWHM of 2.12 mm and an axial FWHM of 13.28 mm (aspect ratio 6.26; Table~\ref{tab:3} and Fig.~\ref{fig:5}b to e). The axial FWHM is about 20\% longer than that of the single-focus lenses at a similar lateral width (11.1 to 12.3 mm and 2.0 to 2.1 mm in S1 and S3; Table~\ref{tab:2}), and the peak gain of 7.59 is comparable, so the focus is lengthened without loss of focal pressure. Stronger elongation for tract-like targets would require a longer target volume in the loss function.

\begin{table}[t]
\centering
\caption{Multiplane volumetric beam-shaping metrics}
\label{tab:3}
\begin{tabular}{@{}lcc@{}}
\toprule
Metric & Cigar focus & Diffraction-limited \\
\midrule
Skull segment                 & S4      & S2      \\
Peak gain ($p/p_{0}$)         & 7.59    & 3.11    \\
Targeting error (mm)          & 0.61    & 0.56    \\
Axial shift (mm)              & $-$0.25 & $-$0.50 \\
Lateral FWHM (mm)             & 2.12    & 2.01 ($1.34\lambda$) \\
Axial FWHM (mm)               & 13.28   & 11.93   \\
Aspect ratio (axial/lateral)  & 6.26    & 5.94    \\
\bottomrule
\end{tabular}

\addtabletext{From the 3D k-Wave validation (fluid model, 1 MHz). Errors are relative to the target at $z = 50$ mm, and lateral and elevational shifts were at or below 0.5 mm.}
\end{table}

For scenarios requiring the preservation of a central structure, such as a calcified pineal gland, a major vessel or the ventricular system, we synthesized a toroidal intensity distribution, which in wave physics corresponds to an acoustic vortex carrying orbital angular momentum. Generating a stable vortex of topological charge $l = 1$ through a heterogeneous skull is difficult, because phase aberration from variable bone thickness tends to destroy the phase singularity and fill the central null. We addressed this by initializing the optimization with a spiral Fresnel warm start, after which the volumetric optimizer fine-tuned the topology for the aberration of segment S2. In k-Wave validation the ring formed at a radius of 9.9 mm, close to the 10 mm design radius, and its core remained darker than the ring from $z = 35$ to 62 mm, with a core-to-ring pressure ratio of 0.35 to 0.51 (Fig.~\ref{fig:5}f to j).

The ring is not uniform: around the ring at $z = 50$ mm the lowest intensity maximum is 0.3 times the highest (Fig.~\ref{fig:5}h). It is also weak relative to the source (peak gain 0.98), because energy is spread over a ring rather than a spot, and the pressure inside the skull reaches 4.4 times the ring peak. Because the validation used a fluid model, this non-uniformity cannot be attributed to shear-mode conversion; it more likely reflects reverberation and strong aberration in segment S2 that the one-way HASA model used for design does not capture. In practice, mode conversion at steep incidence angles could add further azimuthal losses. The vortex result therefore shows that the framework can encode a ring with a darker core through a strongly aberrating segment, while structure-sparing applications would need a more uniform ring and higher ring gain.

A further physical insight from this framework is the distinction between pattern matching and energy maximization when the objective is the tightest possible focal spot. In standard holographic optimization the loss minimizes the error between the simulated field and an idealized target shape such as a 3D Gaussian. A physically perfect focus from a finite aperture, however, carries sidelobes from diffraction at the aperture edges. When the optimizer tries to match a sidelobe-free Gaussian it reads those physical sidelobes as error, and suppresses them by implicitly apodizing the lens, softening transmission at the aperture edge. The beam profile becomes cleaner, but the effective numerical aperture falls, which broadens the central spot and lowers peak pressure.

To avoid this implicit apodization we shifted to energy maximization. Rather than enforcing a beam shape, we defined a sub-wavelength bucket mask at the target coordinates and maximized the intensity integral within it, leaving the algorithm free to distribute energy anywhere outside the mask and so to accept the sidelobes that accompany the tightest central peak while using the full aperture. The optimized topology (Fig.~\ref{fig:5}k) shows high-contrast, Fresnel-like zones consistent with use of the full spatial-frequency bandwidth of the aperture. In k-Wave validation through segment S2 (Table~\ref{tab:3} and Fig.~\ref{fig:5}l to o) the focus formed 0.56 mm from the target with a lateral FWHM of 2.01 mm, or $1.34\lambda$ at 1 MHz, within 15\% of the 1.76 mm pressure FWHM expected for an aberration-free aperture of the same F-number ($1.41\lambda F\#$, $F\# = 0.83$). Compared with the single-focus lens corrected for the same segment (Table~\ref{tab:2}, S2), the energy-maximization lens reduced the targeting error from 1.50 to 0.56 mm and narrowed the focus laterally, from 2.14 to 2.01 mm, and axially, from 13.34 to 11.93 mm, at a 12\% lower peak gain (3.11 against 3.52).

\section{Discussion}

This study shows that modelling wave propagation inside the lens, rather than treating the lens as a thin phase screen, enables high-fidelity holography with 3D-printed lenses at the frequencies at or below 1 MHz used for transcranial ultrasound.

Current rapid design methods for acoustic holograms, such as the IASA, rely on the TEA to generate the lens after optimization \cite{R4}. These models treat the lens as a 2D phase screen and ignore internal wave dynamics. Full-wave time-domain solvers capture that physics but remain computationally prohibitive for iterative clinical-scale design \cite{R8}. Our analysis shows that the TEA breaks down at the sub-megahertz frequencies required for transcranial targeting: near 1 MHz and below, the lens enters an acoustically thick regime in which the high aspect ratio of lens features induces refractive walk-off and diffraction, spreading energy laterally and producing pixel migration and volumetric crosstalk.

HASA-ADAM shifts from phase-screen estimation to true volumetric topology optimization. By optimizing the 3D topology rather than a 2D phase profile it accounts for internal propagation, amplitude modulation and edge diffraction without the burden of time-domain solvers, because the propagation is computed in the frequency domain. This improved PSNR by 3.7 to 5.1 dB over TEA-based designs for the representative holograms in Fig.~\ref{fig:2}b, and by 2.8 dB on average across the 45 configurations of the parametric sweep, while generating large-aperture clinical lenses of about 60 mm in under 20 min with recovered focal confinement and minimized sidelobes.

Beyond that average gain, the regime analysis surfaces two design-relevant insights. First, the advantage of topology optimization is robust across the entire sweep of NA, lens sound speed and frequency: $\Delta$PSNR remains positive in every combination sampled, and persists even at 2.0 MHz, where every design sampled still sits below the $N_\mathrm{w} = 1$ boundary (Fig.~\ref{fig:2}c; Figs.~\ref{fig:S5} to \ref{fig:S7}). Second, $N_\mathrm{w}$ organizes the design space without invoking a material-set thin-to-thick boundary, so each design can be located inside or outside the operating regime defined by $N_\mathrm{w} = 1$. The data also expose a fidelity inversion that runs counter to the usual intuition that higher frequency is better: for topology optimization, SSIM rises as the operating frequency decreases, from 0.63 to 0.70 to 0.76 at 2.0, 1.0 and 0.5 MHz, while that of the TEA lens falls, and for the high-NA design the PSNR at 1.0 MHz matches that at 2.0 MHz despite the doubled wavelength. Together these findings identify the sub-MHz to low-MHz window as a favourable operating regime for transcranial high-fidelity holography, accessible when refractive walk-off is explicitly pre-compensated.

Projecting complex, high-spatial-frequency patterns through a strongly aberrating skull exposes general limitations of acoustic holography in highly heterogeneous media (Fig.~\ref{fig:S13}). The HASA propagator is based on a parabolic, one-way approximation of the heterogeneous Helmholtz equation, which assumes backscattering can be neglected. The large impedance difference between the water coupling medium ($Z_\mathrm{w} \approx 1.5$ MRayl) and dense cortical bone ($Z_\mathrm{b} \approx 5.5$ to 6.0 MRayl) gives a reflection coefficient of about 0.57, so roughly 32\% of the incident energy is reflected. The multi-layered skull also traps sound, producing complex reverberation. Imposing $R = 0$ therefore overestimates transcranial transmission and ignores the phase decorrelation caused by internal standing waves. In addition, the HASA convolution step is derived under the slowly varying envelope approximation, which requires the speed of sound to change slowly relative to the wavelength. At the water-skull boundary the speed of sound jumps from 1500 m/s to over 2500 m/s within a sub-millimeter distance, and this step violates the continuity assumption, introducing spectral leakage and artificial phase accumulation in the spatial Fourier transforms.

Skull geometry may also cause shear conversion. Because of local anatomical curvature, about 50\% of the illuminated skull surface presents an incidence angle above the critical threshold $\theta_\mathrm{c} \approx 30^\circ$ even under perfect registration. Projecting high-spatial-frequency patterns through a strongly aberrating skull therefore imposes a hard boundary beyond which the fluid-based optimizer degrades through multiple reflections, spectral leakage, phase discontinuities and shear errors.

Evaluating a discrete four-point target highlights how focal depth and transcranial propagation interact. In the free field the system resolves the spots effectively at $z = 45$ mm, before the beam widens at deeper planes. Introducing the skull degrades structural definition at all depths: the skull blurs the focus and also restructures the error landscape. While the nominally optimal 45 mm plane retains part of its shape, the compounding effects of near-field blurring and skull-induced phase aberration at 20 mm are highly destructive (Fig.~\ref{fig:S13}). Severe phase distortion from the cranial barrier therefore not only degrades global resolution but uniquely penalizes targets positioned too close to the inner bone interface. These compounded failure modes are worst for complex targets, which is what produces the major drop in correlation of the GT Bee pattern.

In computational cost, HASA-ADAM sits between time-reversal methods \cite{R17,R23} and trained deep-learning frameworks \cite{R24,R25,R26} that are opaque and depend on extensive training data. Methods such as the modified mixed-domain method \cite{R15} offer higher accuracy by including reflections, which HASA currently neglects, but the HASA formulation is highly amenable to automatic differentiation, which is a decisive computational advantage for iterative optimization and gives a favorable balance of speed and accuracy for the forward problem. Optimizing thickness directly is mathematically harder than optimizing phase and takes about twice as long to converge, roughly 15 min at $\lambda/6$ for 500 iterations, and better tuning of hyperparameters and initial conditions should accelerate it further. Hybrid strategies that merge the speed of deep learning with the accuracy and interpretability of HASA propagation on GPUs could further mitigate the $\mathcal{O}(N^{2}\log N)$ scaling and allow the real-time implementations that some applications require \cite{R27,R28}.

Multiplane volumetric optimization further distinguishes the framework from conventional planar holography. Planar optimization is largely blind to the field outside the target slice, which leads to a needle focus with poor axial confinement, or to the apodization trap when matching idealized, sidelobe-free patterns. By computing propagation through the entire region of interest, volumetric gradient descent explicitly constrains energy deposition along the whole beam path. This enables extended cigar foci tailored to elongated anatomical tracts, acoustic vortices with a darker core, and near-diffraction-limited focusing through energy maximization. The volumetric fields also expose a limit of the one-way fluid design model: the vortex ring is non-uniform even in fluid simulations through the most aberrating segment, which points to reverberation and strong aberration in the skull rather than shear-wave generation alone. Capturing and pre-compensating for these 3D elastodynamics will likely be important for precise, arbitrary beam shaping in deep-brain targets.

We envision that the framework could also support noninvasive aberration correction of dense phased arrays \cite{R39}, acting as an objective functional metric for in vivo phase and amplitude optimization of individual elements. Together, these contributions support the design of simple, economical and high-performance ultrasound systems for high-precision neuro-interventions. Such systems may support daily or weekly treatments in outpatient or limited-resource settings without compromising performance, and so support the translation and broad dissemination of this technology \cite{R13,R29}. They may also reduce the heavy reliance on repeated intraoperative MRI during transcranial ultrasound interventions such as targeted drug delivery or liquid biopsy, which can significantly complicate or delay their implementation, given that the average wait for an MRI appointment can be several months \cite{R30}. Complementary ultrasound-based monitoring strategies, including nonlinear acoustic feedback \cite{R41,R42,R45}, parametric-array-mediated transcranial delivery \cite{R43} and high-contrast microbubble imaging at low modulating frequencies \cite{R44}, can be combined naturally with topology-optimized lenses to support real-time treatment guidance.

We also observed discrepancies between simulation and experiment that the compressional model does not explain, most clearly for the small-feature hologram (Fig.~\ref{fig:3}). One possible source is elastic mode conversion at the lens-water interface, which the fluid-based HASA model neglects: steep lens topography produces oblique local incidence angles that can excite slow shear waves in the polymer. A full-wave elastic simulation of the small-feature hologram reduced the mean-squared error relative to the measurement from 0.0230 to 0.0163 and raised the correlation from 0.864 to 0.942 (Fig.~\ref{fig:S11}). This single comparison is indicative rather than conclusive, and registration, material-property uncertainty and printing tolerances may also contribute. A systematic study of mode conversion in holographic lenses is a direction for future work.

Several limitations point the way forward. First, the fidelity of topology optimization is sensitive to the material properties of the printed lens, and sub-wavelength discrepancies in sound speed from manufacturing and curing variation can shift the focus axially, so post-print calibration is required. Second, clinical scenarios involving trapped postoperative gas or large variations in bone porosity induce severe impedance mismatches ($Z_\mathrm{skull}/Z_\mathrm{air} \approx 7500$) that scatter primary energy and destroy phase coherence. Third, the HASA propagator used for iterative design relies on a fluid model. Although our simulations suggest that shear modes in the skull, and possibly in the lens, affect focal fidelity, incorporating full elastic propagation into a topology optimization loop remains computationally expensive for rapid clinical design and prototyping. Future versions should integrate computationally efficient elastic approximations, which could pioneer shear-mode holography in which passive lenses are deliberately crafted to exploit mode conversion to minimize off-target skull heating or enhance transcranial energy transfer.

By accelerating the design of high-fidelity holographic lenses, this work could support the clinical translation and broad dissemination of transcranial holography. While our focus is biomedical, high-fidelity acoustic holography is relevant to a much broader range of applications \cite{R4,R8,R31,R32,R33}, and our results highlight the importance of optimizing the full lens topology whenever field fidelity is critical.

\section{Methods}
\subsection{Construction of the transcranial medium}

Patient-specific transcranial optimization requires a realistic heterogeneous 3D map of the acoustic properties of the skull. The anatomical geometry was acquired by micro-CT. Because the ex vivo skull remained secured in its fixture throughout the scan, the DICOM volumes were co-registered directly with the transducer coordinate space. The raw 3D data were interpolated onto a uniform Cartesian grid at an isotropic voxel resolution of \SI{250}{\micro\meter} ($\sim\lambda/6$) to satisfy the 1 MHz sampling requirement. CT voxel intensities were normalized against water and air baselines to obtain approximate Hounsfield units $H$, which were used to compute the volumetric porosity fraction of the bone, $\Psi = 1 - H/1000$. A porosity-dependent semi-empirical mixture model \cite{R34,R35} then converted these values to acoustic properties: the density $\rho$ and longitudinal speed of sound $c$ of each voxel were linearly interpolated between pure water ($c = 1480$ m/s, $\rho = 1000$ kg/m$^{3}$) and dense cortical bone ($c_\mathrm{max} = 2500$ m/s, $\rho_\mathrm{max} = 2200$ kg/m$^{3}$), in proportion to the solid bone fraction $1 - \Psi$.

\subsection{Heterogeneous wave propagation}

For a time-harmonic pressure field $\tilde{p}(\boldsymbol{r})e^{-i\omega t}$ at angular frequency $\omega$, the angular spectrum $P$ is the 2D spatial Fourier transform

\begin{equation}
P(k_x, k_y, z) = \mathcal{F}_k[\tilde{p}(x,y,z)].
\label{eq:3}
\end{equation}

\noindent In heterogeneous media where the spatial variation of the sound speed $c(\boldsymbol{r})$ is small compared with the wavelength, the ordinary differential equation for $P$ becomes

\begin{equation}
P_{z} + k_{z}^{2}P = \Lambda * P,
\label{eq:4}
\end{equation}

\noindent where $\Lambda = \mathcal{F}_k[k_0^2(1-\mu)]$, $k_0 = \omega/c_0$, $\mu = c_0^2/c^2$, $c_0$ is a reference sound speed, and $*$ denotes 2D convolution over $k_x$ and $k_y$. The skull and tissue speed-of-sound and density maps came from micro-CT of a human skull \cite{R34,R35}, at an original resolution of \SI{95}{\micro\meter} binned to \SI{150}{\micro\meter}, which is 10 points per wavelength at $f_0 = 1$ MHz for $c_0 = 1480$ m/s in water. HASA is designed to handle such spatially varying properties, and this model captures refraction and transmission through the skull. Absorption was not included in the present implementation.

An implicit solution of Eq.~\ref{eq:4} follows from a Green's function technique \cite{R36}, and numerical approximation gives $P$ at arbitrary $z$ through

\begin{equation}
P^{n+1} \approx P^{n}e^{i k_{z}\Delta z}
+ \frac{e^{i k_{z}\Delta z}}{2 i k_{z}}\left(P^{n} * \Lambda\right)\Delta z,
\label{eq:5}
\end{equation}

\noindent where $P^{n} = P(k_x, k_y, n\Delta z)$. Provided the marching step $\Delta z$ is much smaller than the wavelength, Eq.~\ref{eq:5} computes the field in the heterogeneous medium with numerical stability. We used $\Delta z = \SI{250}{\micro\meter}$, that is $\lambda/6$. With this propagator the target-plane pressure $P_t$ follows efficiently from the initial field $P_0$.

\subsection{Optimization parameters}

To adapt the optimization across frequencies at a fixed aperture of 60 mm and focal depth of 45 mm, the spatial sampling relative to the wavelength was held constant, with voxel sizes $dx$ and $dz$ kept at $\lambda/6$ to prevent numerical aliasing. The maximum lens thickness $t_\mathrm{max}$ sets the ability to impart a full $2\pi$ phase shift; because the phase delay is proportional to frequency, higher frequencies require thinner lenses for the same wrapping (Table~\ref{tab:4}).

\begin{table}[b]
\centering
\caption{Summary of optimization parameters}
\label{tab:4}
\begin{tabular}{@{}lcccc@{}}
\toprule
$f$ & $\lambda$ & $dx$ & $N$ & $t_\mathrm{max}$ \\
\midrule
0.5 MHz & 3.00 mm & 0.500 mm & 256 & 10.0 mm \\
1.0 MHz & 1.50 mm & 0.250 mm & 240 & 5.0 mm  \\
2.0 MHz & 0.75 mm & 0.125 mm & 512 & 2.5 mm  \\
\bottomrule
\end{tabular}

\addtabletext{Grid size $N$ is the number of lateral samples; $t_\mathrm{max}$ is the maximum lens thickness.}
\end{table}

\subsection{Approach 1: topology optimization}

Topology optimization refines the material thickness distribution directly to achieve a desired intensity profile at the target plane, using HASA to model skull heterogeneity and propagation through complex media. The algorithm initializes with a uniform thickness distribution $u^{(0)}$ and iteratively minimizes the loss. At iteration $n$ the thickness is

\begin{equation}
t^{(n)} = \min\left\{\mathrm{softplus}\left(u^{(n)}\right), t_\mathrm{max}\right\},
\label{eq:6}
\end{equation}

\noindent where softplus enforces non-negative thickness and $t_\mathrm{max} = 5$ mm bounds the hologram. The complex field $P^{(n)}$ follows from the HASA forward model $\mathcal{H}(P_0, t^{(n)})$, where $P_0$ is the initial pressure distribution of the transducer. The loss quantifies the mismatch between the computed intensity $I^{(n)} = |P^{(n)}|^{2}$ and the target intensity $I_t$ at each location:

\begin{equation}
\mathcal{L}^{(n)} = \sum_{x,y}\left|I^{(n)}(x,y) - I_t(x,y)\right| + \lambda\mathcal{R}\left(t^{(n)}\right),
\label{eq:7}
\end{equation}

\noindent where $\lambda\mathcal{R}(t^{(n)})$ encourages smooth thickness variations with $\lambda = 0.01$ setting the regularization strength. ADAM with learning rate $\eta = 0.01$ updates the thickness parameter:

\begin{equation}
u^{(n+1)} = u^{(n)} - \eta\nabla_{u}\mathcal{L}^{(n)}.
\label{eq:8}
\end{equation}

\noindent Gradients are computed by automatic differentiation in TensorFlow. Iteration continues until the loss falls below $\epsilon = 0.001$ or the maximum iteration count $N_\mathrm{max}$ is reached, 500 for single-point focusing and 2000 for complex 2D distributions (Table~\ref{tab:1}).

\subsection{Approach 2: phase optimization}

Phase optimization refines the phase distribution $\phi$ at the hologram surface while holding the amplitude profile fixed. It begins from a uniform phase $\phi^{(0)}$ and zero thickness, and at each iteration computes

\begin{equation}
P_t^{(n)} = \mathcal{H}\left(P_0, \phi^{(n)}\right),
\label{eq:9}
\end{equation}

\noindent where $\mathcal{H}$ is the HASA propagation operator, with target-plane intensity $I_t^{(n)} = |P_t^{(n)}|^{2}$ and loss

\begin{equation}
\mathcal{L}^{(n)} = \sum_{x,y}\left|I_t^{(n)} - I_t\right| + \lambda\mathcal{R}\left(\phi^{(n)}\right),
\label{eq:10}
\end{equation}

\noindent in which $\mathcal{R}(\phi^{(n)})$ regularizes the phase profile toward smoothness. The phase is updated by gradient descent with ADAM. After convergence the optimized phase is converted to a thickness map, and the routine accounts for the effect of hologram thickness on amplitude transmission. The conversion uses

\begin{equation}
\Delta\phi(x,y) = \left(k_{w} - k_{h}\right)\Delta h(x,y),
\label{eq:11}
\end{equation}

\noindent where $k_w$ and $k_h$ are the wavenumbers in water and in the hologram material. The thickness in turn gives the transmission coefficient $\alpha_T$ and the complex amplitude at the hologram plane, following established expressions \cite{R4} (Table~\ref{tab:1}).

\subsection{Volumetric and multiplane optimization}

Planar optimization minimizes $\mathcal{L}_\mathrm{planar} = \||P(x,y,z_\mathrm{target})| - T(x,y)\|_{1}$, defined only at the target slice, which can produce the needle effect. We instead implemented volumetric gradient descent, using a recurrent computational graph to record the complex field history $U_k$ at every slice $z_k$ within the 3D region of interest $\Omega$. The volumetric loss is

\begin{equation}
\mathcal{L}_\mathrm{vol} = \sum_{k=z_\mathrm{start}}^{z_\mathrm{end}} w_{k}
\left\|\frac{|U_{k}|^{2}}{\max\left(|U_{k}|^{2}\right)} - T_\mathrm{vol}^{(k)}\right\|_{1}
+ \lambda_\mathrm{reg}\|\nabla\mathcal{T}\|^{2},
\label{eq:12}
\end{equation}

\noindent where $T_\mathrm{vol}^{(k)}$ is the 3D target volume at slice $k$, $w_k$ allows axial weighting, and the final term regularizes the lens topology $\mathcal{T}$.

For elongated deep-brain targets we solved the inverse source problem with an anisotropic 3D Gaussian target

\begin{equation}
T_\mathrm{cigar}(x,y,z) = \exp\left(-\frac{x^{2}+y^{2}}{2\sigma_\mathrm{xy}^{2}}
- \frac{(z-z_{c})^{2}}{2\sigma_{z}^{2}}\right),
\label{eq:13}
\end{equation}

\noindent with $\sigma_z \gg \sigma_\mathrm{xy}$. For the acoustic vortex ($l = 1$) the optimization was initialized with a spiral Fresnel warm start, which supplies the spiral phase delay needed to steer rays onto a ring of the desired radius; the volumetric optimizer then fine-tuned the topology to pre-compensate for the aberration of the specific skull segment.

For diffraction-limited focusing, and to circumvent the apodization trap of pattern matching, we used an energy-maximization formulation with a sub-wavelength mask $M(\boldsymbol{r})$ of radius $\sigma < \lambda$ centred on the target:

\begin{equation}
\mathcal{L}_\mathrm{energy} = -\iiint_{\Omega}|P(\boldsymbol{r})|^{2}M(\boldsymbol{r})\,\mathrm{d}V.
\label{eq:14}
\end{equation}

\noindent Minimizing $\mathcal{L}_\mathrm{energy}$ leaves the algorithm unrestricted outside the mask, allowing the constructive sidelobes required to use the full aperture.

\subsection{Fabrication}

For topology optimization the thickness mask is 3D-printed without further processing. For phase optimization, Eq.~\ref{eq:11} converts the optimized phase map to a thickness map, which is then printed. We used Clear White v4 resin (Formlabs, Somerville, MA) on a Formlabs Form 3B printer. This resin has low attenuation across the frequency range of interest and a relatively high speed of sound, with group velocity $c_g = 2591$ m/s and attenuation $\alpha_0 = 2.922$ dB MHz$^{-1.044}$ cm$^{-1}$, which makes it well suited to 3D-printed acoustic holograms among the materials characterized by Bakaric et al. \cite{R37}.

\subsection{Trans-skull simulations}

Three-dimensional linear acoustic simulations used the open-source k-Wave toolbox \cite{R38} with GPU acceleration. Elastic effects were ignored because the angle of incidence on the skull bone was below the critical angle in most cases. All simulations used a spatial grid of $\Delta x = \Delta y = \SI{250}{\micro\meter}$ and a CFL number below 0.3 for numerical stability. A porosity-dependent semi-empirical relationship converted Hounsfield units from the micro-CT scan into medium properties, and spatially varying maps of sound speed, density and attenuation were used throughout. The maximum speed and density of solid bone were taken as 2500 m/s and 2000 kg/m$^{3}$. Spatially dependent attenuation was assigned by a power law $\alpha = \alpha_0 f^{\beta}$ with uniform exponent $\beta = 1.2$ (Table~\ref{tab:5}).

\begin{table}[b]
\centering
\caption{Acoustic parameters for the k-Wave fluid and elastic simulations}
\label{tab:5}
\setlength{\tabcolsep}{2pt}
\begin{tabular}{@{}lcc@{}}
\toprule
Parameter & Fluid model & Elastic model \\
\midrule
Background density $\rho_{0}$        & 1000 kg/m$^{3}$ & 1000 kg/m$^{3}$ \\
Density threshold $\rho_\mathrm{bone}$ & $>$1100 kg/m$^{3}$ & $>$1100 kg/m$^{3}$ \\
Compressional speed $c_{p}$          & from CT & from CT \\
Shear speed $c_{s}$                  & n/a     & 1400 m/s \\
Fluid attenuation $\alpha_{0}$       & 0.0022 dB/cm/MHz$^{y}$ & 0.0022 dB/cm/MHz$^{y}$ \\
Compressional atten.\ $\alpha_{p}$   & from CT & from CT \\
Shear attenuation $\alpha_{s}$       & n/a     & 10 to 20 dB/cm/MHz$^{y}$ \\
Courant number (CFL)                 & 0.3     & 0.1 \\
\bottomrule
\end{tabular}
\end{table}

For the trans-skull validation of the optimized lenses (Fig.~\ref{fig:4}i and j, Fig.~\ref{fig:5}, Tables~\ref{tab:2} and \ref{tab:3}), the lens and skull-segment property maps were resampled to a 0.25 mm grid, 240 $\times$ 240 voxels laterally and 6 points per wavelength in water at 1 MHz, and the fields were computed with the GPU solver for a continuous 1 MHz source of 0.25 MPa over a 60 mm circular aperture in the first $z$ plane, with a time step of 20 ns and the maximum pressure recorded throughout the volume. The lens ($c = 2591$ m/s, $\rho = 1178$ kg/m$^{3}$) was voxelized from the thickness map and stacked directly in front of the skull segment, so it occupies as many $z$ slices as its maximum thickness requires. During optimization the lens was modelled as a fixed 5 mm block, whereas the corrected lenses have a maximum thickness of 3.1 to 3.7 mm, so in the validation runs the skull sat 1.25 to 1.75 mm closer to the source than in the design model, while the free-field lenses at 5 mm were validated in the design geometry. This mismatch acts against the corrected lenses, so the reported improvements are likely conservative. The corrected lenses for segments S1 to S3 were designed with the volumetric objective using an elongated Gaussian target, whereas the uncorrected baseline is the single-plane free-field lens; this difference in objective may contribute to the slightly longer axial FWHM of the corrected foci (Table~\ref{tab:2}). Focal metrics were computed from the maximum-pressure volume: the focal peak was taken beyond $z = 30$ mm, past the skull; targeting errors are relative to the target at $z = 50$ mm from the source plane, which is 45 mm beyond a 5 mm lens; FWHM values are full widths at half the peak pressure within $\pm$10 mm of the peak; the focal volume is the $-$3 dB main lobe containing the peak; and the skull volume is the number of skull voxels ($c > 1700$ m/s) at or above $-$3 dB of the focal peak.

To capture shear mode conversion and propagation at the lens interfaces, we also ran full-wave 3D elastic simulations with the \texttt{pstdElastic3D} solver in k-Wave. Skull bone was isolated by a density threshold ($\rho > 1100$ kg/m$^{3}$), with a shear wave speed of 1400 m/s and shear attenuation of 20 dB/cm assigned to those voxels. Because sharp fluid-solid boundaries can trigger numerical instability in elastic simulations, we smoothed the property matrices and held the CFL number to 0.1.

\subsection{Trans-skull experiments}

Ex vivo trans-skull experiments were performed in a degassed deionized water tank. A 60 mm diameter piston transducer (Precision Acoustics, Dorchester, UK) was coupled to a 3D-printed hologram lens and driven by a function generator (33511B; Keysight) through a power amplifier (E\&I, Rochester, NY). The transducer and lens assembly was attached to the parietal region of a skull cap (Skulls Unlimited, Oklahoma City, OK) that had been degassed overnight for about 12 h. The focal field was scanned with a calibrated 2 mm needle hydrophone (Precision Acoustics) on a three-axis positioning system (Velmex, Bloomfield, NY) and recorded on a digital oscilloscope (Pico Technology, St Neots, UK).

\subsection{Forward modelling of hydrophone aperture and sampling}

Blurring and pixelation differences between simulation and measurement were reproduced with a forward model of the acquisition chain. First, the simulated field $P_\mathrm{sim}(x,y)$, generated on the \SI{250}{\micro\meter} grid, was convolved with a spatial kernel representing the active area of the needle hydrophone (active element diameter $D = 2.0$ mm):

\begin{equation}
P_\mathrm{avg}(x,y) = P_\mathrm{sim}(x,y) * H_\mathrm{disk}(D).
\label{eq:15}
\end{equation}

\noindent Performing this on the full-resolution grid captures the spatial averaging over the sensor face. Second, the digitisation of the stepper-motor scan was modelled by downsampling $P_\mathrm{avg}$ to the experimental scan grid, which had a \SI{500}{\micro\meter} step:

\begin{equation}
P_\mathrm{fwd}(i,j) = P_\mathrm{avg}(x_i, y_j)
\quad \text{for } x,y \in \mathrm{Grid}_\mathrm{exp}.
\label{eq:16}
\end{equation}

\noindent The result is a synthetic measurement that reproduces both the aperture blur and the step-limited pixelation seen in the data.

\subsection{Free-field measurements and fidelity metrics}

Free-field fields of the large and small GT Bee lenses (1 MHz, designed for a 45 mm projection distance) were raster-scanned in degassed water with a 0.2 mm needle hydrophone over a 121 $\times$ 121 grid: 60 $\times$ 60 mm at 0.5 mm step for the large lens, 30 $\times$ 30 mm at 0.25 mm step for the small one, with 16 averages per point. The per-point acoustic energy of the recorded waveform was used as the measured intensity, which tracks the squared amplitude of the 1 MHz band-passed waveform ($r = 0.99$ and 0.89 for the large and small scans). The expected field of each printed lens was computed from the thickness map used to generate its print file, with the HASA forward model through the voxelized lens (0.25 mm grid; $c = 2591$ m/s and $\rho = 1178$ kg/m$^{3}$ in the resin; 60 mm uniform source), followed by angular-spectrum propagation through water to the design plane. On a lens design simulated with both models, this prediction agreed with k-Wave at $r = 0.84$ on the design plane. The predicted intensity and the design target were interpolated onto the scan grid and registered to the measurement by the in-plane translation ($\pm$8 mm, 0.25 mm steps) that maximized the Pearson correlation with the measured map; no rotation, flip or rescaling was applied. Each map was normalized to its own peak, and PSNR (unit data range) and SSIM (11 $\times$ 11 Gaussian window, $\sigma = 1.5$) were computed with the same convention as the simulation study. The Pearson correlation coefficient $r$ is also reported.

\pdfbookmark[0]{Data and code availability}{bm:avail}
\section*{Data and code availability}
All data needed to evaluate the conclusions of this study are present in the paper and Supplementary Information. The code for the HASA-ADAM framework will be made available upon publication.

{\small
\bibliographystyle{naturemag}
\bibliography{references}
}

\pdfbookmark[0]{Acknowledgements}{bm:ack}
\section*{Acknowledgements}
We thank Hohyun Lee for the hydrophone scans of the large GT Bee hologram (Fig.~\ref{fig:3}, \textit{top right}), Dr.~Stas Emelianov for providing the E\&I high-power amplifier, and Drs.~S.~Schoen Jr.~and L.~Degertekin, as well as the LEAP HI team, for their valuable input, especially during the early phase of this research. This study was supported by National Science Foundation grants 1933158 and 1830577 (P.P.D., C.D.A.), from the Division of Civil, Mechanical and Manufacturing Innovation, Leading Engineering for America's Prosperity, Health, and Infrastructure.

\pdfbookmark[0]{Author contributions}{bm:contrib}
\section*{Author contributions}
P.P.D. and C.D.A. designed research; P.P.D. performed research and analyzed data; C.D.A. supervised the project; P.P.D. and C.D.A. wrote the paper.

\pdfbookmark[0]{Competing interests}{bm:coi}
\section*{Competing interests}
The authors declare no competing interest.

\clearpage
\onecolumn
\setcounter{section}{0}\setcounter{figure}{0}\setcounter{table}{0}\setcounter{equation}{0}
\renewcommand{\thesection}{S\arabic{section}}
\renewcommand{\thefigure}{S\arabic{figure}}
\renewcommand{\thetable}{S\arabic{table}}
\renewcommand{\theequation}{S\arabic{equation}}
\pdfbookmark[0]{Supplementary Information}{bm:si}
\begin{center}
{\sffamily\bfseries\fontsize{16}{20}\selectfont\color{accent} Supplementary Information\par}
\vspace{7pt}
{\large Differentiable topology optimization for high-fidelity acoustic holography beyond the thin-element approximation\par}
\vspace{5pt}
{Pradosh P. Dash and Costas D. Arvanitis\par}
\vspace{3pt}
{\footnotesize Correspondence: costas.arvanitis@gatech.edu\par}
\vspace{7pt}
{\color{accent}\rule{\textwidth}{1.2pt}}
\end{center}
\vspace{3pt}
\noindent{\footnotesize This file contains Supplementary Sections S1 to S11, Figs. S1 to S14,
Tables S1 and S2, and legends for Movies S1 and S2.\par}
\vspace{8pt}

\section{Optimizer hyperparameters and computational cost}

\begin{figure}[htbp]
\centering
\includegraphics[width=5.2in]{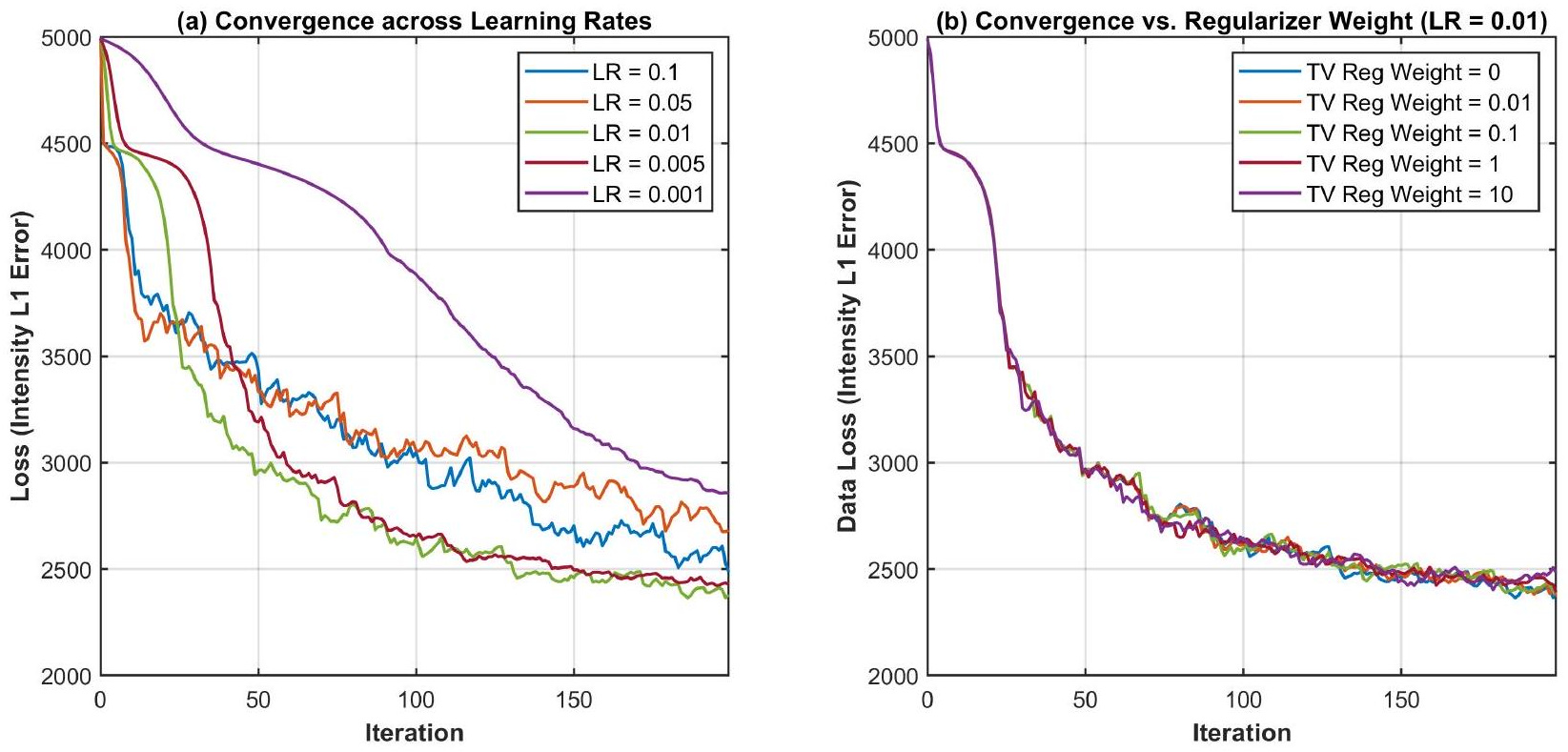}
\caption{\textbf{HASA-ADAM hyperparameter study.} (\textit{a}) Optimization convergence across learning rates. (\textit{b}) Pure data loss across total-variation regularizer weights. The coincident plateaus confirm that the regularizer shapes the topology without reducing focal fidelity.}
\label{fig:S1}
\end{figure}

Stochastic optimizers can be sensitive to hyperparameter choices, so we ran an empirical sweep to check convergence across configurations. Sweeping the learning rate over $\eta \in \{0.001, 0.005, 0.01, 0.05, 0.1\}$ gave stable, monotonic loss curves in every case (Fig.~\ref{fig:S1}\textit{a}). High learning rates fell fast but oscillated near convergence, and low rates were stable but slow, so $\eta = 0.01$ was the best practical compromise. We also checked the total-variation regularizer. Total-variation smoothing is needed for manufacturability, but an excessive weight could compromise acoustic accuracy. Sweeping $\lambda \in \{0.0, 0.01, 0.1, 1.0, 10.0\}$ while tracking the raw intensity $L_1$ error over 200 iterations showed no separation between the curves (Fig.~\ref{fig:S1}\textit{b}), which confirms that the regularizer shapes the lens geometry without degrading the acoustic reconstruction. The full set of values used throughout the study is in Table~\ref{tab:S1}.

\begin{figure}[htbp]
\centering
\includegraphics[width=4.0in]{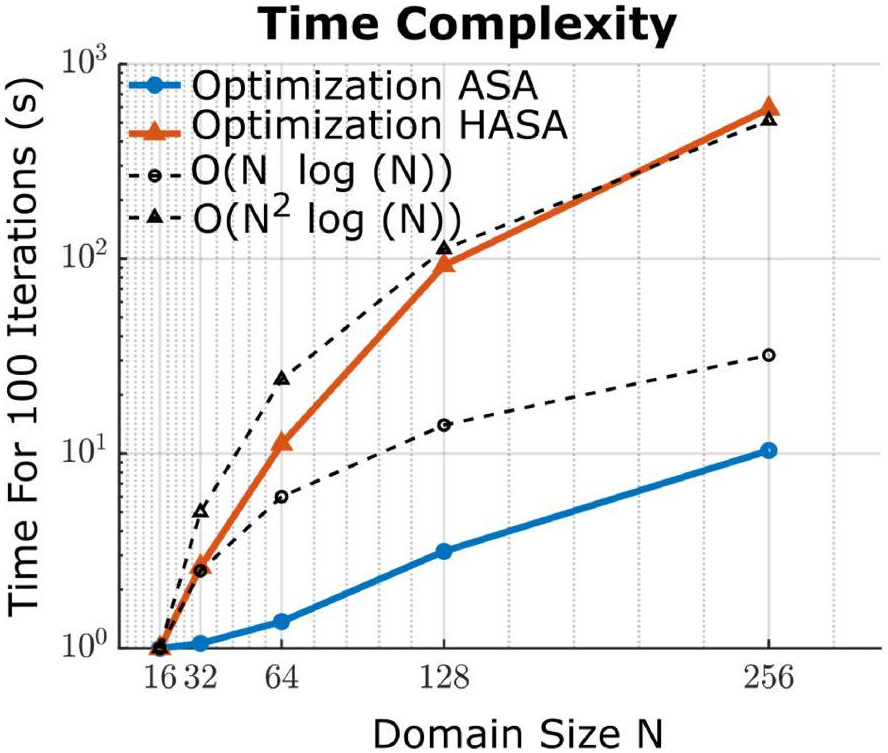}
\caption{\textbf{Time complexity.} HASA optimization, $\mathcal{O}(N^{2}\log N)$, requires more computational time than the homogeneous angular spectrum approach, $\mathcal{O}(N\log N)$.}
\label{fig:S2}
\end{figure}

The cost of the heterogeneous propagator relative to the homogeneous angular spectrum approach is shown in Fig.~\ref{fig:S2}: HASA scales as $\mathcal{O}(N^{2}\log N)$ against $\mathcal{O}(N\log N)$, which is the price of carrying an in-plane varying speed-of-sound map through every marching step.

\begin{table}[htbp]
\centering
\caption{HASA-ADAM hyperparameters for topology optimization}
\label{tab:S1}
\begin{tabular}{@{}ll@{}}
\toprule
Parameter & Value \\
\midrule
Learning rate ($\eta$)              & 0.01 \\
Maximum iterations ($N_\mathrm{max}$) & 500 \\
Optimizer                           & ADAM \\
Initialization ($T_{0}$)            & 3.6 mm \\
Loss function metric                & $L_1$ intensity loss \\
Variable constraint                 & $\min(\mathrm{softplus}(u),\ 5\ \mathrm{mm})$ \\
Spatial padding                     & $N/2$ on all boundaries \\
\bottomrule
\end{tabular}
\end{table}

\section{Material properties and model sensitivity}

\subsection*{Lens sound speed measured by through-transmission}

The longitudinal speed of sound in the Clear White v4 resin was measured by a through-transmission substitution method. A reference signal was acquired through a water path, followed by five measurements with a 3D-printed resin sample of thickness $d = 15$ mm inserted into the acoustic path at different positions between the transducer and the hydrophone. Arrival times were determined by peak-threshold detection on the signal envelope obtained from the Hilbert transform, $E(t) = |\mathcal{H}(s(t))|$, and the time shift $\Delta t$ was taken as the difference between the sample and baseline arrival times. The sample sound speed follows from the relative time of flight, given $c_\mathrm{water} \approx 1480$ m/s:

\begin{equation}
c_\mathrm{sample} = \left(\frac{1}{c_\mathrm{water}} - \frac{\Delta t}{d}\right)^{-1}.
\label{eq:S1}
\end{equation}

\begin{figure}[htbp]
\centering
\includegraphics[width=4.6in]{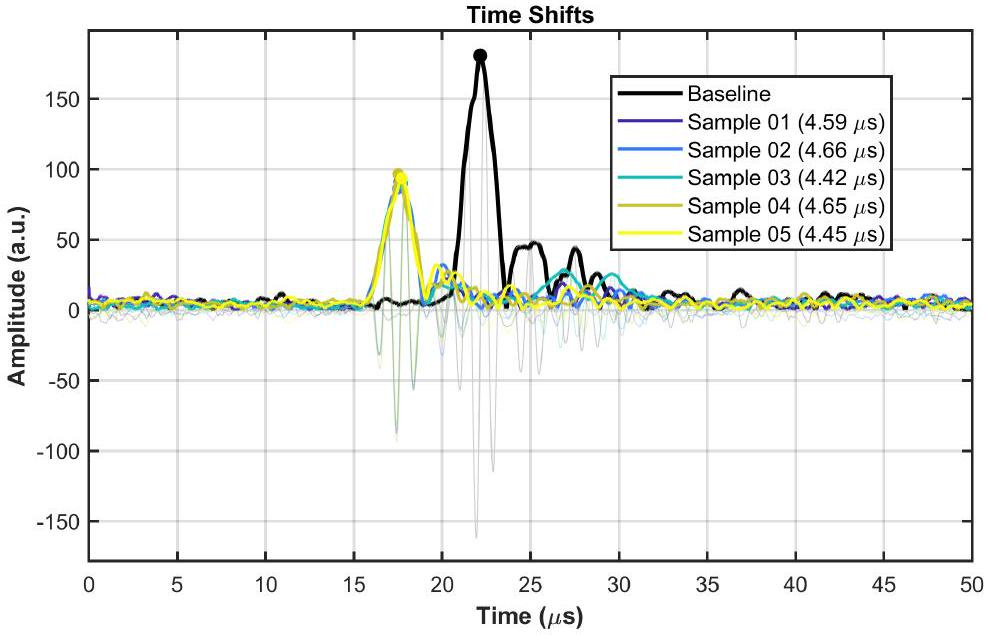}
\caption{\textbf{Speed-of-sound estimation by through-transmission measurement.}}
\label{fig:S3}
\end{figure}

The five trials gave time shifts of 4.5920, 4.6640, 4.4240, 4.6480 and 4.4480 \si{\micro\second}, a mean of $\Delta t = 4.56 \pm 0.11$ \si{\micro\second} and therefore $c_\mathrm{exp} = 2689 \pm 54$ m/s (Fig.~\ref{fig:S3}). This is within about 3.8\% of the value used in design, $c_\mathrm{design} = 2590$ m/s, and well within the spread expected from curing variation.

\subsection*{Lens sound-speed mismatch produces a longitudinal shift}

Let $c_\mathrm{des}$ be the sound speed assumed during optimization and $c_\mathrm{real}$ the speed of the lens material after curing. The phase accumulated through a thickness $h$ is governed by the refractive contrast with the background medium $c_{0}$:

\begin{equation}
\phi_\mathrm{real}(x,y) = \omega h(x,y)\left(\frac{1}{c_{0}} - \frac{1}{c_\mathrm{real}}\right).
\label{eq:S2}
\end{equation}

Because the optimizer computed $h(x,y)$ from $c_\mathrm{des}$, substituting back shows that the realized phase is the target phase scaled by a constant mismatch ratio $\gamma$:

\begin{equation}
\phi_\mathrm{real}(x,y) = \phi_\mathrm{target}(x,y) \times
\underbrace{\left[\frac{1/c_{0} - 1/c_\mathrm{real}}{1/c_{0} - 1/c_\mathrm{des}}\right]}_{\gamma}.
\label{eq:S3}
\end{equation}

For single-point focusing, a lens sound-speed error therefore shifts the hologram along $z$ rather than destroying spatial coherence. In the paraxial approximation a focusing element imparts a quadratic phase profile, $\phi_\mathrm{target}(r) \approx -k r^{2}/2F_\mathrm{des}$, so the mismatch ratio simply describes a perfect lens of new focal length $F_\mathrm{new} = F_\mathrm{des}/\gamma$: internal phase relationships are preserved and the plane of image formation is displaced axially.

\subsection*{Skull sound-speed and frequency variation}

Errors in estimating the sound speed of the highly heterogeneous skull bone translate into uncertainty in the aberration correction. A deviation of $\pm$15\%, $c = c_\mathrm{design}(1 \pm 0.15)$, gives a one-way travel-time error across a 7 mm skull of about 2.64 rad (151$^\circ$) at 1 MHz.

Varying the skull sound speed by $\pm$15\% in 3D simulations for two focusing configurations ($F\#0.75$ and $F\#1.0$) produced roughly a $\pm$30\% variation in peak focal pressure, with axial drift peaking at 2.0 mm, slightly larger than one acoustic wavelength in water at 1 MHz. Running the transducer off resonance scales the beam width and depth of field inversely with frequency, so that a higher frequency tightens and attenuates the beam while a lower one broadens and deepens it, moving the focus to $z(f) = (f_{0}/f)z_{0}$; $\pm$5\% frequency shifts produce axial shifts comparable to two or three wavelengths. The statistics for both sweeps are collected in Table~\ref{tab:S2}.

\begin{table}[htbp]
\centering
\caption{Peak amplitude and location statistics for sound-speed and frequency errors}
\label{tab:S2}
\begin{tabular}{@{}lccccc@{}}
\toprule
Configuration & Min & Max & Range & Mean & S.D. \\
\midrule
\multicolumn{6}{@{}l}{\textit{Sound-speed mismatch ($\pm$15\%): peak amplitude (normalized)}} \\
$F\#0.75$ (45 mm) & 0.737 & 1.352 & 0.615 & 1.016 & 0.200 \\
$F\#1.00$ (60 mm) & 0.735 & 1.299 & 0.564 & 1.027 & 0.197 \\
\addlinespace
\multicolumn{6}{@{}l}{\textit{Sound-speed mismatch ($\pm$15\%): peak axial location shift (mm)}} \\
$F\#0.75$ (45 mm) & $-$0.250 & 1.750 & 2.000 & 0.500 & 0.685 \\
$F\#1.00$ (60 mm) & $-$1.250 & 0.500 & 1.750 & $-$0.212 & 0.548 \\
\addlinespace
\multicolumn{6}{@{}l}{\textit{Frequency mismatch ($\pm$5\%): peak amplitude (normalized)}} \\
$F\#0.75$ (45 mm) & 0.850 & 1.173 & 0.323 & 1.003 & 0.112 \\
$F\#1.00$ (60 mm) & 0.859 & 1.282 & 0.423 & 1.059 & 0.147 \\
\addlinespace
\multicolumn{6}{@{}l}{\textit{Frequency mismatch ($\pm$5\%): peak axial location shift (mm)}} \\
$F\#0.75$ (45 mm) & $-$2.750 & 4.500 & 7.250 & 0.591 & 2.432 \\
$F\#1.00$ (60 mm) & $-$3.250 & 3.250 & 6.500 & 0.045 & 2.159 \\
\bottomrule
\end{tabular}
\end{table}

\section{Comparison with IASA phase optimization}

\begin{figure}[htbp]
\centering
\includegraphics[width=6.0in]{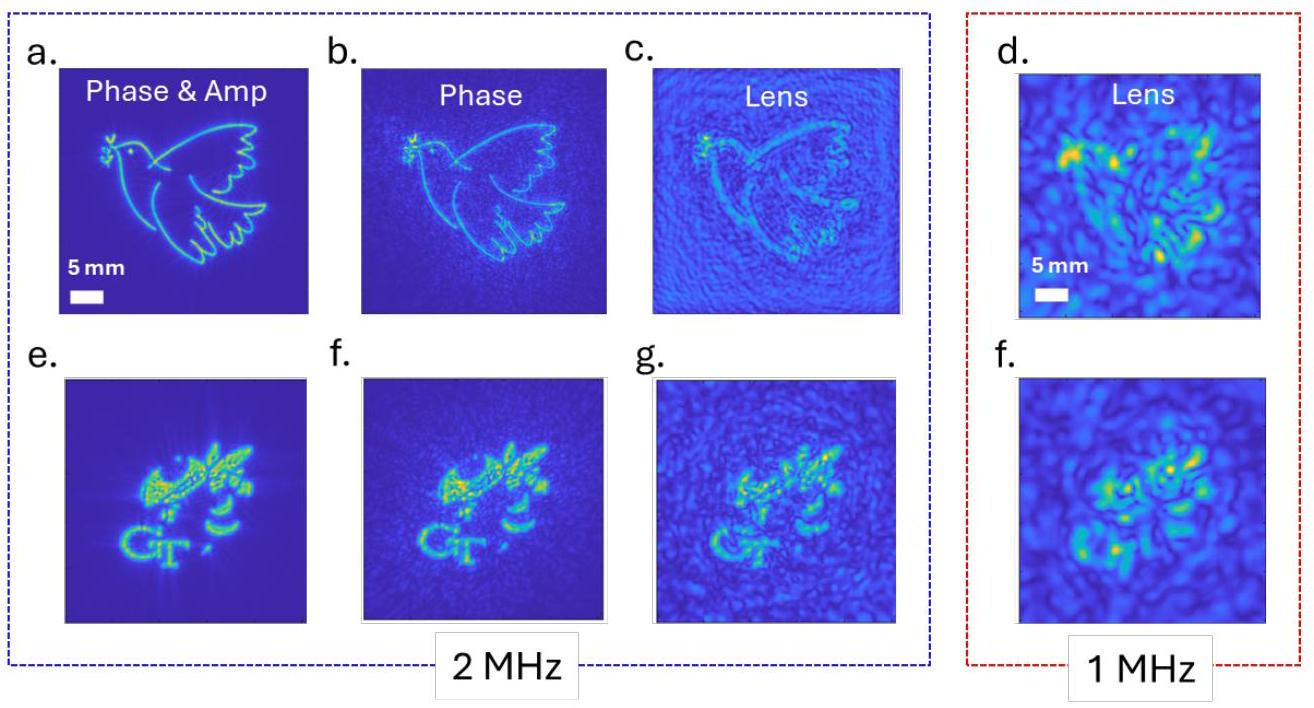}
\caption{\textbf{Effect of frequency scaling on hologram fidelity.} (\textit{Blue box}) 2 MHz regime, comparing reconstruction of the dove (\textit{top row}) and the Georgia Tech logo (\textit{bottom row}): (\textit{a}, \textit{e}) ideal phase and amplitude; (\textit{b}, \textit{f}) phase-only (IASA); (\textit{c}, \textit{g}) physical lens. At 2 MHz the IASA phase map is coherent, but the physical lens shows distortion from refractive walk-off. (\textit{Red box}) 1 MHz regime: scaling the logo design to 1 MHz collapses the fidelity, because the diffraction limit $\delta$ doubles and merges the fine details, and the IASA-designed lens fails to form a coherent image.}
\label{fig:S4}
\end{figure}

To locate the failure of the thin-element approximation against an established method, we re-examined benchmark holograms from Melde et al.~\cite{R4} and the Georgia Tech logo under conventional iterative angular spectrum approach (IASA) phase optimization, and scaled them in frequency (Fig.~\ref{fig:S4}).

At 2 MHz ($\lambda \approx 0.75$ mm) the diffraction limit is fine enough to resolve both targets, and the phase-only approximation produces coherent images for the dove and the logo alike. The physical lens, however, introduces distortion: the lens is acoustically thick, and the resulting refractive walk-off is something IASA cannot predict.

Scaling to 1 MHz at the same aperture ($D = 50$ mm) and distance ($z = 20$ mm) brings the system to the diffraction limit. Halving the frequency doubles the minimum feature size, so the fine lines of the logo fall below the acoustic point-spread function, and standard IASA optimization fails: it attempts to create features that the physics cannot support, and returns noisy, unrecognizable distributions.

Comparing the two methods under the geometric constraints used throughout this study ($f = 1$ MHz, $D = 60$ mm, $z = 45$ mm) gives the same result. As the scaling analysis predicts, IASA failed to converge to a valid solution: the resolution limit ($\delta \approx 1.13$ mm) combined with the volumetric thickness of the lens produces incoherent scattering. HASA-ADAM topology optimization pre-compensated for the distortion and produced a coherent reconstruction of the same target.

\section{Diffraction and thickness scaling set the regime}

Lowering the frequency is not by itself a solution, because resolution degrades with it. For a hologram to resolve a target feature of lateral size $\delta$ at a distance $z$, the system parameters must satisfy

\begin{equation}
\underbrace{2\,\mathrm{d}x}_{\text{sampling limit}} \leq
\underbrace{\frac{c}{f}}_{\lambda} \lesssim
\underbrace{\delta}_{\text{resolution}} \approx
\underbrace{\frac{c\,z}{f D}}_{\text{diffraction geometry}}.
\label{eq:S4}
\end{equation}

Resolution is therefore inversely proportional to frequency, and reducing the frequency linearly expands the focal spot, blurring fine detail. Under the geometric constraints of this study ($f = 1$ MHz, $D = 60$ mm, $z = 45$ mm) the diffraction limit is $\delta \approx 1.13$ mm, comparable to the finest target features, so the system is diffraction-constrained.

The thin-element approximation breaks down because the lens thickness does not scale down as favourably as the feature size. Unlike an electronic phased array, a passive lens generates its phase delay by propagation through a material of differing sound speed, so a full $2\pi$ of wrapping requires a modulation depth set by the sound-speed contrast:

\begin{equation}
L_{2\pi} = \frac{\lambda}{\left|1 - c_{0}/c_\mathrm{lens}\right|}.
\label{eq:S5}
\end{equation}

This is the full-wrap height $H_{2\pi}$ of the main text, written with the sound speeds rather than with $n = c_\mathrm{lens}/c_{0}$: $L_{2\pi} = \lambda/(1 - c_{0}/c_\mathrm{lens}) = n\lambda/(n-1)$. For biocompatible polymers in water this contrast is moderate, about 0.6, which requires 3 to 5 mm of thickness at 1 MHz. At megahertz frequencies the total lens thickness is therefore often much larger than the wavelength.

\section{Full reconstruction sweeps across NA and lens sound speed}

\begin{figure}[htbp]
\centering
\includegraphics[width=6.5in]{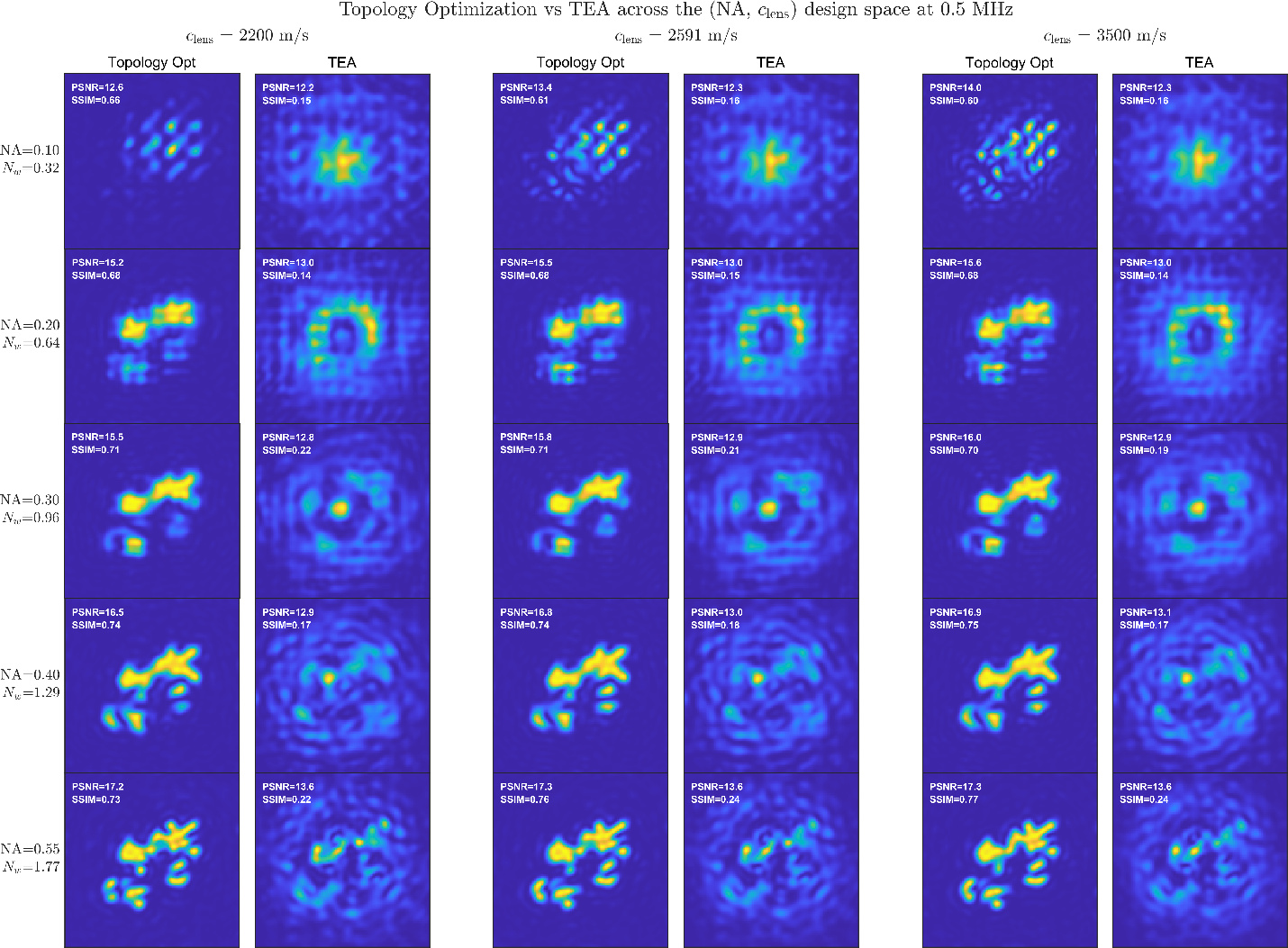}
\caption{\textbf{Reconstruction sweep at 0.5 MHz on the GT Bee target.} Rows are the five sampled numerical apertures ($\mathrm{NA} = 0.10$, 0.20, 0.30, 0.40 and 0.55, top to bottom). Three column pairs show $c_\mathrm{lens} = 2200$, 2591 and 3500 m/s; within each pair the left panel is the topology-optimization reconstruction and the right is the thin-element physical-lens reconstruction. PSNR (dB) and SSIM are overlaid in white at the top left of each panel.}
\label{fig:S5}
\end{figure}

\begin{figure}[htbp]
\centering
\includegraphics[width=6.5in]{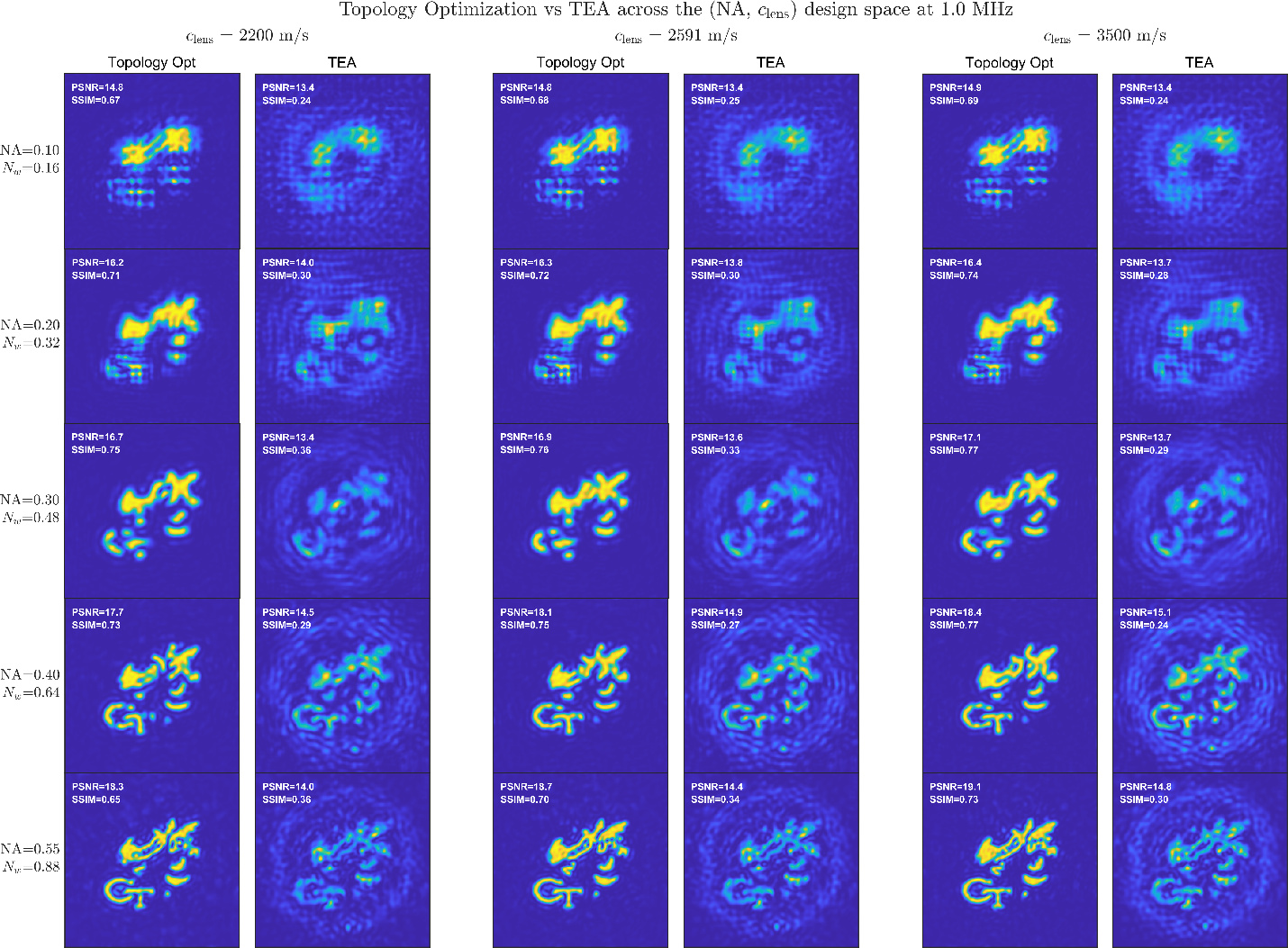}
\caption{\textbf{Reconstruction sweep at 1.0 MHz.} Layout as in Fig.~\ref{fig:S5}. This is the design frequency closest to the regime boundary ($N_\mathrm{w} \approx 0.3$ to 1.9 across the sweep), where topology optimization is expected to separate most clearly from the thin-element approximation.}
\label{fig:S6}
\end{figure}

\begin{figure}[htbp]
\centering
\includegraphics[width=6.5in]{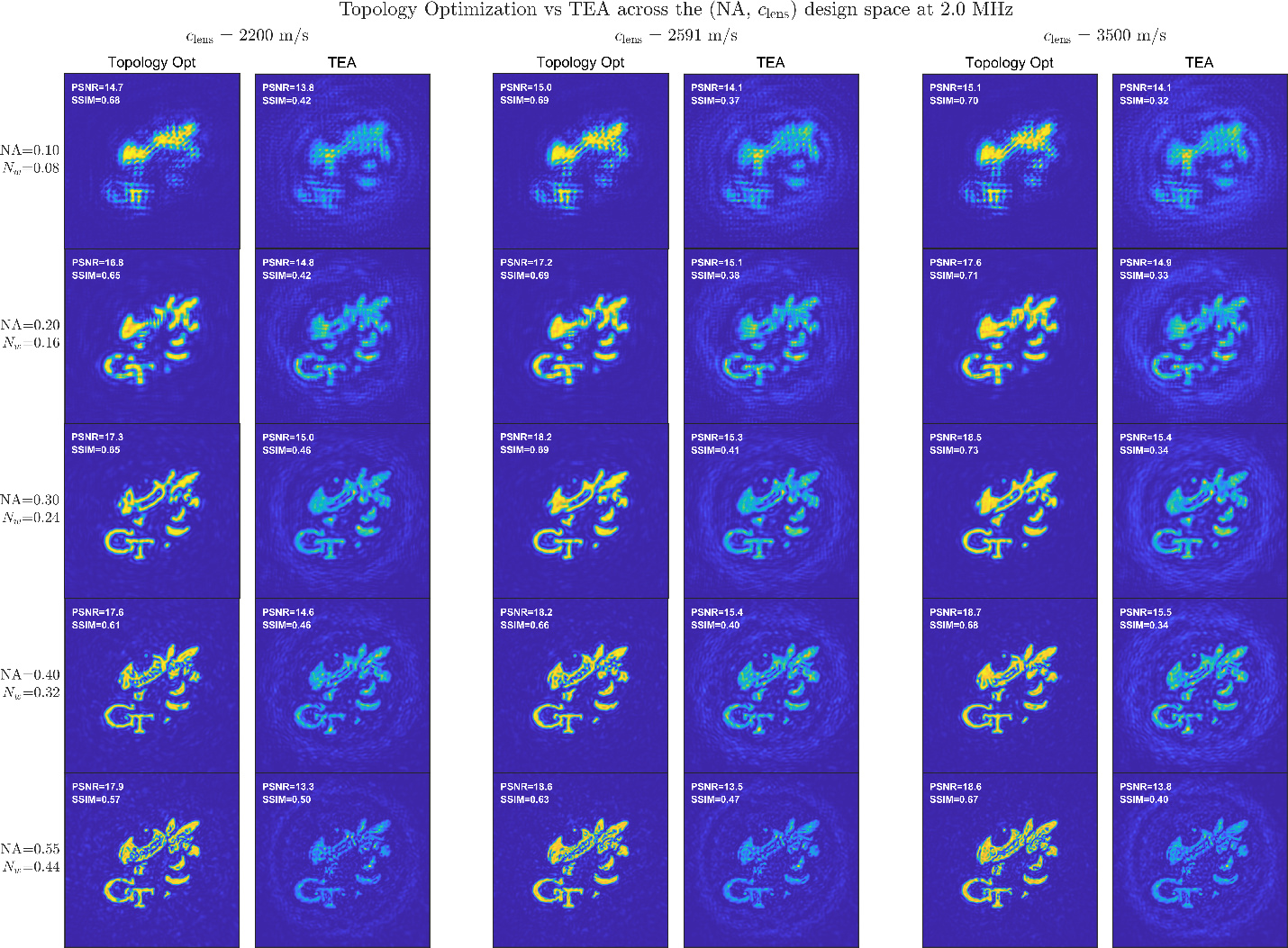}
\caption{\textbf{Reconstruction sweep at 2.0 MHz.} Layout as in Fig.~\ref{fig:S5}. Even at 2.0 MHz, nominally well inside the regime where the thin-element approximation is sufficient, topology optimization keeps a measurable PSNR and SSIM advantage in every (NA, $c_\mathrm{lens}$) combination, because of edge diffraction and localized scattering at the dense phase-wrap discontinuities of the thin-element lens.}
\label{fig:S7}
\end{figure}

\begin{figure}[htbp]
\centering
\includegraphics[width=6.8in]{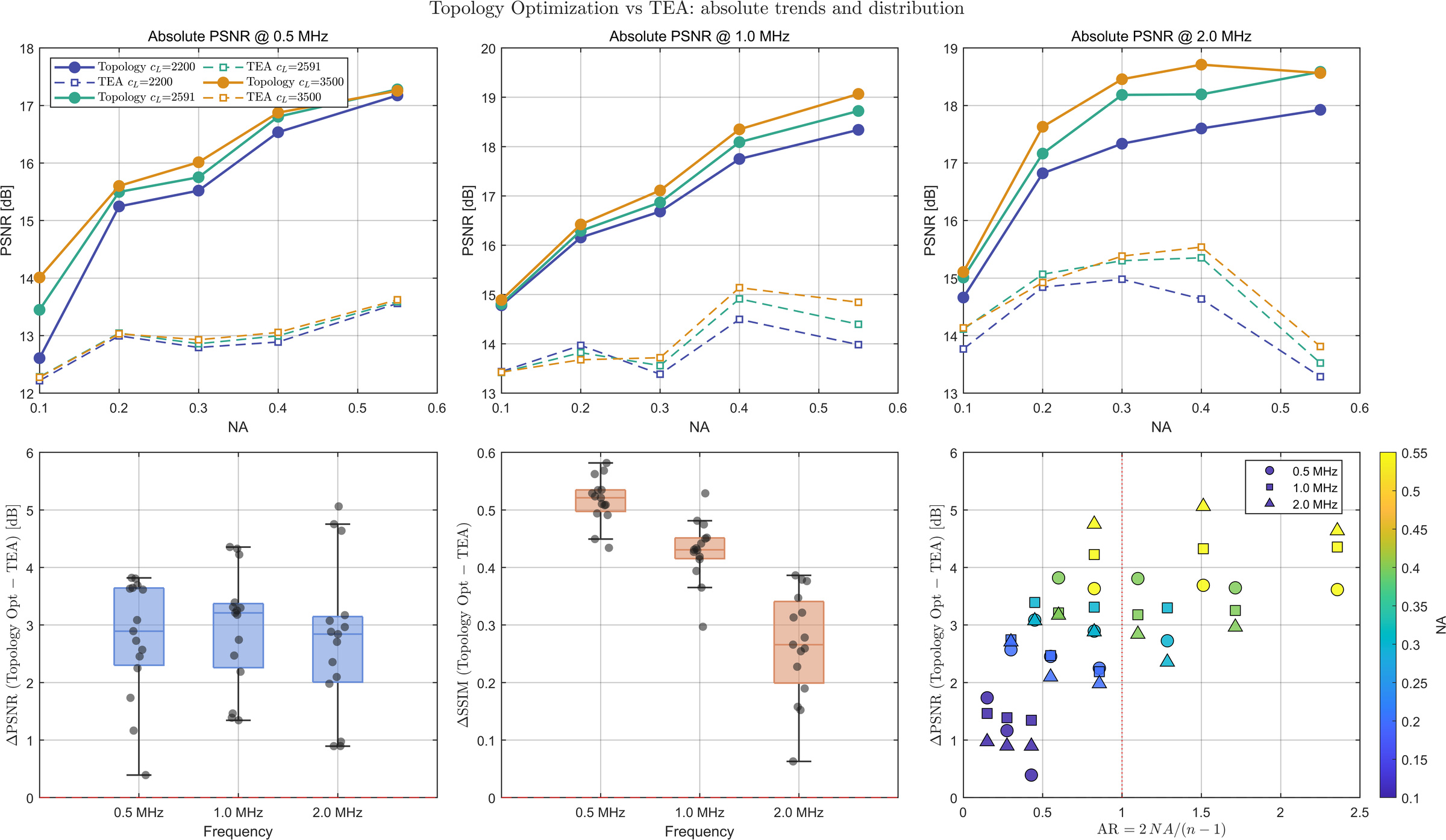}
\caption{\textbf{Effect of lens sound speed on the topology-optimization advantage.} \textit{Top}: absolute PSNR versus NA at 0.5, 1.0 and 2.0 MHz for topology optimization (solid, filled) and the thin-element approximation (dashed, open), for $c_\mathrm{lens} = 2200$, 2591 and 3500 m/s. \textit{Bottom}: distributions of $\Delta$PSNR and $\Delta$SSIM (topology optimization minus thin-element) by frequency, and $\Delta$PSNR versus $2\,\mathrm{NA}/(n-1)$, coloured by NA. That last quantity is the aspect ratio $\mathrm{AR} = 2n\,\mathrm{NA}/(n-1)$ divided by $n$, so it rescales the axis within each lens sound speed without changing the trend. The three sound speeds give nearly overlapping trends, which is why Fig.~2\textit{c} of the main text reports their mean $\pm$ s.d.}
\label{fig:S8}
\end{figure}

Figures~\ref{fig:S5} to \ref{fig:S7} give the complete reconstruction sweeps behind Fig.~2\textit{c} of the main text, at 0.5, 1.0 and 2.0 MHz. Figure~\ref{fig:S8} separates the three lens sound speeds, whose effect on the comparison is minor, which is what justifies averaging over them in the main text.

\section{Spatial-frequency content and the paraxial limit}

The thin-element approximation inherits the paraxial approximation, in which the axial wavenumber is expanded to first order on the assumption that the transverse spatial frequencies are small relative to $k$:

\begin{equation}
k_{z} = \sqrt{k^{2} - \left(k_{x}^{2} + k_{y}^{2}\right)}
\approx k - \frac{k_{x}^{2} + k_{y}^{2}}{2k}.
\label{eq:S6}
\end{equation}

\begin{figure}[htbp]
\centering
\includegraphics[width=6.0in]{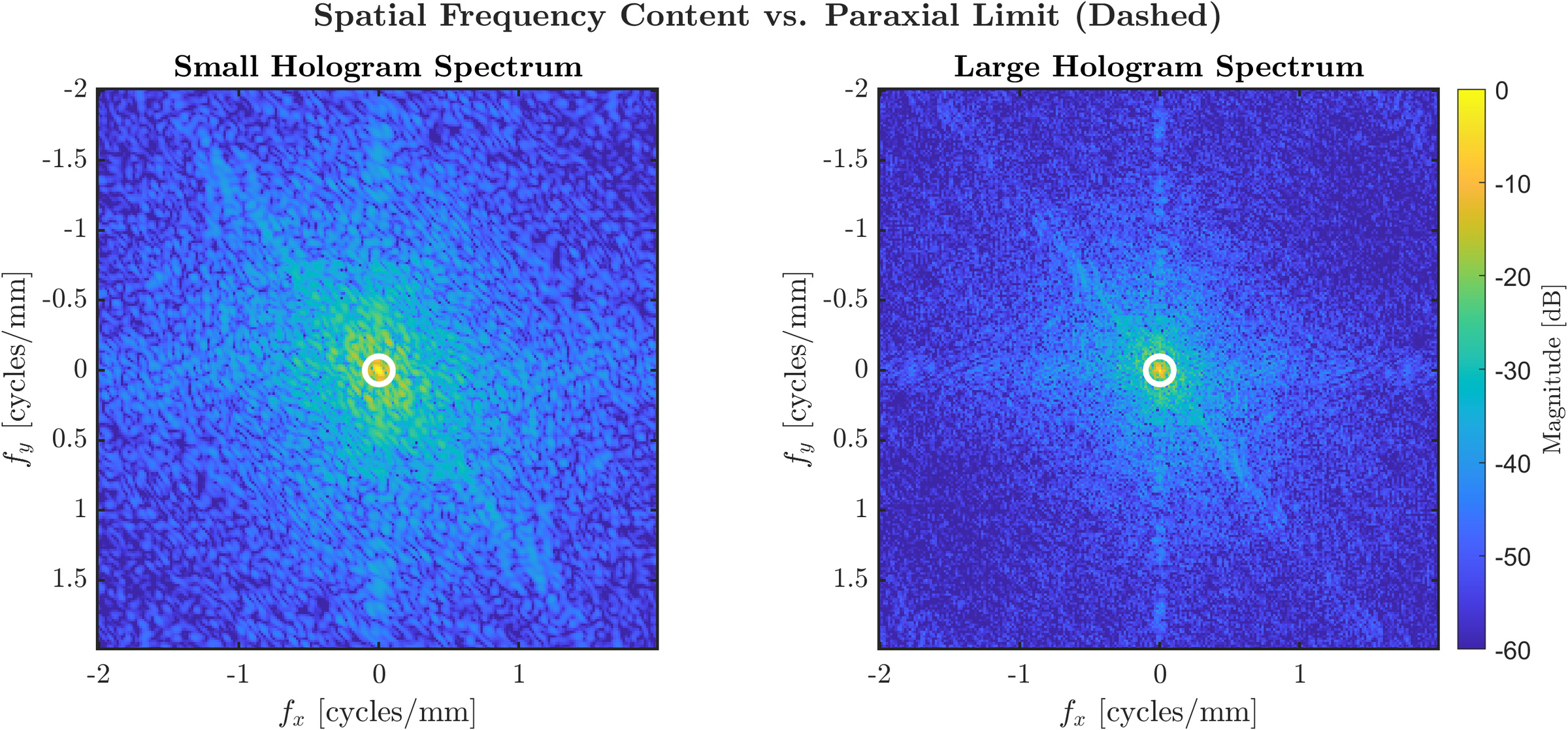}
\caption{\textbf{Spatial-frequency spectra of the small and large target holograms.} The dashed white circle is the theoretical limit of the paraxial approximation. The large target confines its structural energy to lower spatial frequencies and so adheres more closely to paraxial wave mechanics.}
\label{fig:S9}
\end{figure}

This explains why the thin-element approximation performs better on the large hologram than on the small one, even though the absolute image quality of the large target is limited by aperture truncation at short propagation distances. Compressing the target features, as in the small hologram, requires steep diffraction angles, which shifts the structural energy of the target into high transverse spatial frequencies where $(k_x^2 + k_y^2) \ll k^2$ is severely violated and the phase errors become catastrophic. Scaling the target up broadens the physical features and concentrates the energy at lower spatial frequencies. Figure~\ref{fig:S9} shows this directly: the spectral energy of the large hologram falls largely within the paraxial bound, so the thin-element approximation retains phase coherence.

Quantitatively, the absolute performance gap between topology optimization and the thin-element approximation is much smaller for the large target, averaging $\Delta\mathrm{SSIM} \approx 0.15$, than for the small one, $\Delta\mathrm{SSIM} \approx 0.48$. Expressed as retained fidelity, the thin-element approximation holds up to 66.7\% of the structural fidelity achieved by topology optimization for large features (59.5\%, 66.7\% and 54.1\% at 30, 45 and 60 mm), but collapses to between 48.7\% and 30.1\% for the fine features over the same distances. While the absolute image quality drops for large targets because of finite numerical aperture, the mechanics of the thin-element approximation are substantially more robust for low-spatial-frequency features.

\section{Assumptions of the one-way heterogeneous propagator}

HASA marches the field forward under the parabolic, or one-way, approximation of the wave equation, and its spectral convolution step relies in addition on a slowly varying envelope. Both assumptions have consequences at the water-skull interface, which is why they are stated here explicitly.

Propagation in a heterogeneous, non-absorbing medium is governed by the Helmholtz equation, $\nabla^{2}P + k^{2}(\boldsymbol{r})P = 0$, where $k(\boldsymbol{r}) = \omega/c(\boldsymbol{r})$. Factoring the field into a slowly varying envelope on a carrier along the propagation axis, $P(x,y,z) = \psi(x,y,z)e^{ik_{0}z}$ with $k_{0} = \omega/c_{0}$, and substituting gives the exact envelope equation

\begin{equation}
\nabla_{\perp}^{2}\psi + \frac{\partial^{2}\psi}{\partial z^{2}}
+ 2ik_{0}\frac{\partial\psi}{\partial z}
+ \left(k^{2}(\boldsymbol{r}) - k_{0}^{2}\right)\psi = 0,
\label{eq:S7}
\end{equation}

\noindent with $\nabla_{\perp}^{2} = \partial^{2}/\partial x^{2} + \partial^{2}/\partial y^{2}$. The parabolic approximation requires the envelope to evolve slowly along the axis relative to the wavelength, $|\partial^{2}\psi/\partial z^{2}| \ll |2k_{0}\,\partial\psi/\partial z|$, and dropping the second-order axial derivative leaves

\begin{equation}
2ik_{0}\frac{\partial\psi}{\partial z}
= -\nabla_{\perp}^{2}\psi - \left(k^{2}(\boldsymbol{r}) - k_{0}^{2}\right)\psi.
\label{eq:S8}
\end{equation}

A second-order differential equation in $z$ supports both forward and backward solutions; reducing it to first order admits only the forward one. The formulation therefore assumes 100\% forward transmission across every interface. At the water-to-skull boundary, where the reflection coefficient is about 0.57, the algorithm enforces $R = 0$ and overestimates transcranial transmission, and the optimizer is blind to the phase decorrelation caused by internal standing waves and backscattering within the diploe layer.

The spectral convolution relies separately on a Wentzel-Kramers-Brillouin treatment, in which the amplitude $A(z)$ and the accumulated phase $\phi(z)$ of $P(z) = A(z)e^{i\phi(z)}$ are decoupled. Substituting this ansatz into the one-dimensional Helmholtz equation and isolating the real part gives an exact expression for the local phase gradient,

\begin{equation}
\left(\frac{\mathrm{d}\phi}{\mathrm{d}z}\right)^{2} = k^{2}(z) + \frac{1}{A}\frac{\mathrm{d}^{2}A}{\mathrm{d}z^{2}},
\label{eq:S9}
\end{equation}

\noindent which reduces to $\mathrm{d}\phi/\mathrm{d}z \approx \pm k(z)$ only when $|A^{-1}\mathrm{d}^{2}A/\mathrm{d}z^{2}| \ll k^{2}(z)$, that is, when the medium is stable over a wavelength. Substituting $k = \omega/c$ and generalizing to three dimensions, the requirement becomes

\begin{equation}
\frac{1}{k_{0}}\left|\frac{\nabla c}{c}\right| \ll 1.
\label{eq:S10}
\end{equation}

At the water-skull interface the sound speed jumps from about 1480 m/s to over 2500 m/s across a fraction of a wavelength. That step violates the inequality and breaks the decoupled-phase assumption, so propagating waves accumulate artificial phase and produce spectral leakage in the spatial Fourier transforms.

\section{Elastic mode conversion at the lens interface}

The experimental fields in the main text show a diffuse background haze at small hologram feature sizes that the compressional model does not reproduce. Here we examine whether elastic mode conversion at the lens-water interface can account for that residual discrepancy. The analysis is indicative rather than conclusive, and is presented as a mechanism to be explored further.

\begin{figure}[htbp]
\centering
\includegraphics[width=5.6in]{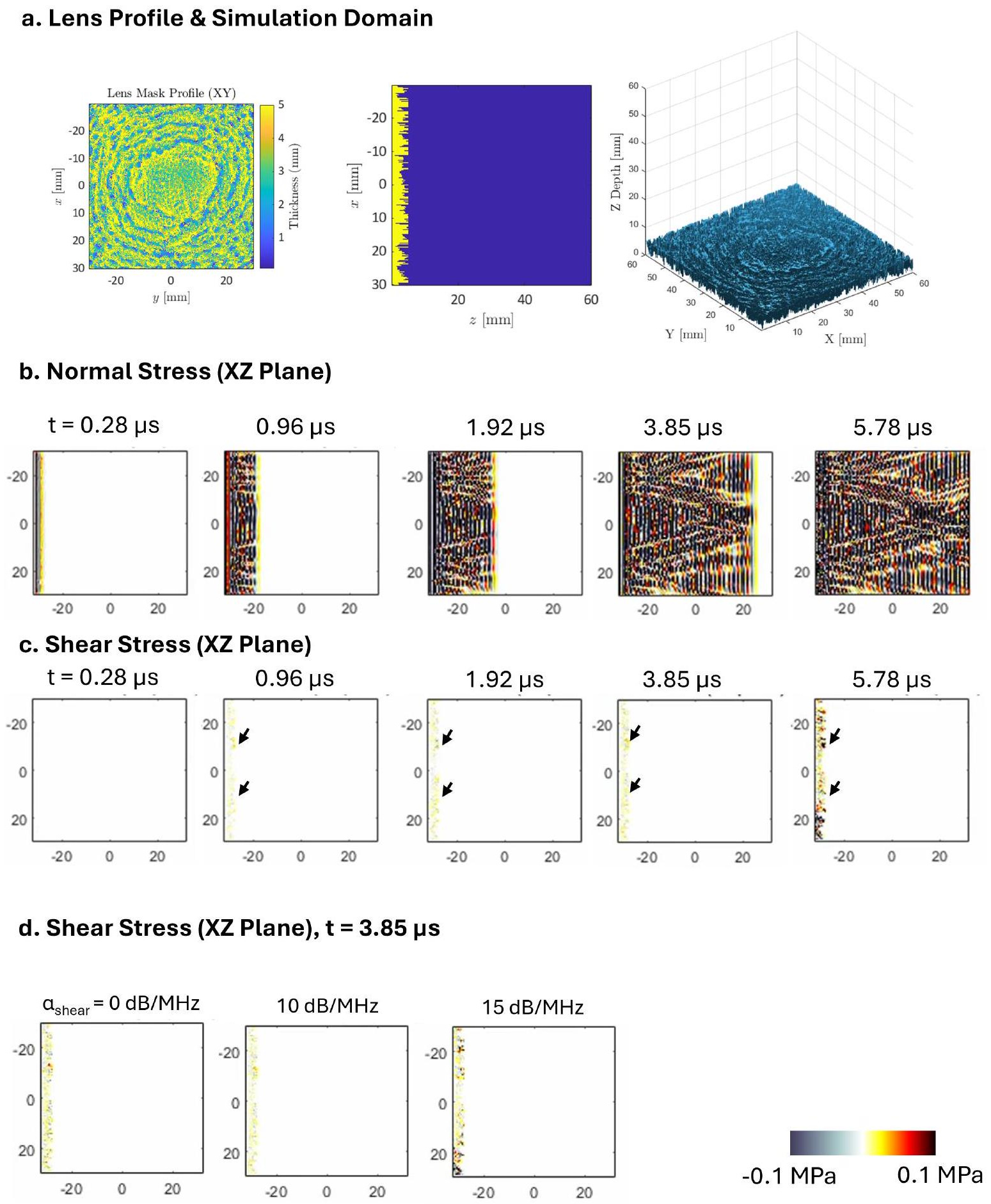}
\caption{\textbf{Elastic wave propagation through the lens topology.} (\textit{a}) Acoustic lens topology and simulation domain. (\textit{b}) Temporal evolution of normal stress in the $xz$ plane. (\textit{c}) Temporal evolution of shear stress in the $xz$ plane; shear waves are generated at the solid-fluid interface, with black arrows marking the origin points. (\textit{d}) Shear stress distribution in the $xz$ plane at a fixed time for shear attenuation coefficients of 0, 10 and 15 dB/MHz.}
\label{fig:S10}
\end{figure}

We first analysed the normal and shear stress evolution at the lens interface with the k-Wave elastic solver, assuming a shear attenuation of 10 dB/MHz within the lens. The solver integrates the coupled first-order velocity-stress equations, $\rho\,\partial v_{i}/\partial t = \partial\sigma_{ij}/\partial x_{j}$ and $\partial\sigma_{ij}/\partial t = \lambda\delta_{ij}\partial v_{k}/\partial x_{k} + \mu(\partial v_{i}/\partial x_{j} + \partial v_{j}/\partial x_{i})$, where $v_i$ is particle velocity, $\sigma_{ij}$ the stress tensor, and $\lambda$ and $\mu$ the Lam\'{e} parameters. The compressional and shear speeds follow as $c_{p} = \sqrt{(\lambda + 2\mu)/\rho}$ and $c_{s} = \sqrt{\mu/\rho}$, and because $\lambda$ and $\mu$ are positive for solids, $c_{p} > c_{s}$. As expected, the normal stress propagated rapidly through the lens while the shear wave lagged, given its lower speed of about 1300 m/s, and the shear contribution was accentuated at the steep lens-water interface (Fig.~\ref{fig:S10}). Shear waves vanished beyond the interface, because water cannot support them. Raising the shear attenuation from 0 to 10 and then 15 dB/MHz progressively reduced the shear stress amplitude; these values are far above the compressional attenuation of a resin-based printed lens, about 3 dB/MHz \cite{R37}, and are used here only as worst cases.

The partitioning of energy at a fluid-solid boundary is governed by the Zoeppritz equations: an incident compressional wave arriving at a non-normal angle partitions into a transmitted compressional wave and a mode-converted shear wave. For the photopolymer used here the shear velocity, about 1300 m/s, is roughly half the longitudinal velocity of 2590 m/s, and the slower modes degrade holographic fidelity in two ways. They accumulate significant phase delay relative to the primary longitudinal wavefront, so that on exiting the lens this energy radiates with uncorrelated phase and interferes destructively, reducing peak focal intensity. They are also refracted at steeper angles, following Snell's law for elastic media,

\begin{equation}
\frac{\sin\theta_{s}}{c_{s}} = \frac{\sin\theta_{p}}{c_{p}},
\label{eq:S11}
\end{equation}

\noindent which steers energy away from the designated target features and disperses it across the observation plane as the diffuse haze seen in the measurements.

\begin{figure}[htbp]
\centering
\includegraphics[width=6.2in]{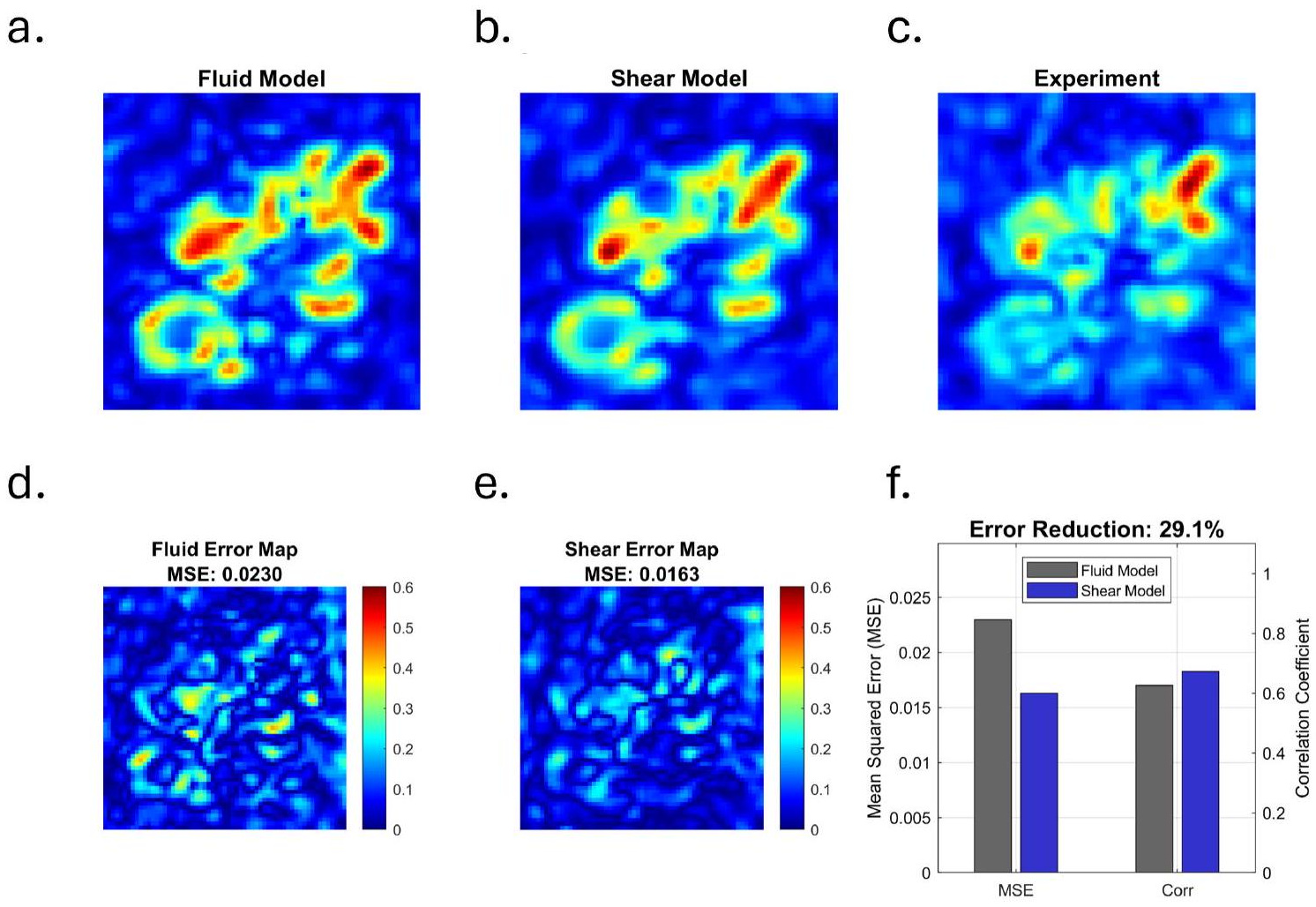}
\caption{\textbf{Compressional, elastic and measured fields for the small-feature hologram.} (\textit{a} to \textit{c}) Normalized focal-plane intensity for the fluid model, the elastic model and the measurement. (\textit{d}, \textit{e}) Absolute error maps relative to the measurement. (\textit{f}) Mean-squared error and spatial correlation. Including mode conversion reduces the mean-squared error from 0.0230 to 0.0163, a 29.1\% reduction, and raises the correlation from $R = 0.864$ to $R = 0.942$. These data indicate a possible contribution from elastic mode conversion and are not presented as conclusive proof of the mechanism.}
\label{fig:S11}
\end{figure}

Including these effects in a full-wave elastic simulation reproduces the morphological distortions and background scattering of the measurement (Fig.~\ref{fig:S11}). The fluid model predicts a confined, high-contrast focal distribution and leaves prominent structural residuals, with a mean-squared error of 0.0230 against the measurement, whereas the elastic model reduces the error to 0.0163, a 29.1\% reduction, and raises the spatial correlation from $R = 0.864$ to $R = 0.942$. To confirm that the smearing follows from slow-shear-wave aberration rather than simple attenuation, we repeated the elastic simulation with the shear absorption set to zero while keeping $c_{s} = 1400$ m/s; the resulting field was virtually identical to the fully attenuating case (correlation 1.00, SSIM 0.99, negligible mean-squared error), which places the origin of the blurring in phase aberration and refraction at the steep lens interfaces rather than in energy absorption.

\section{Operating window and failure modes}

\begin{figure}[htbp]
\centering
\includegraphics[width=3.6in]{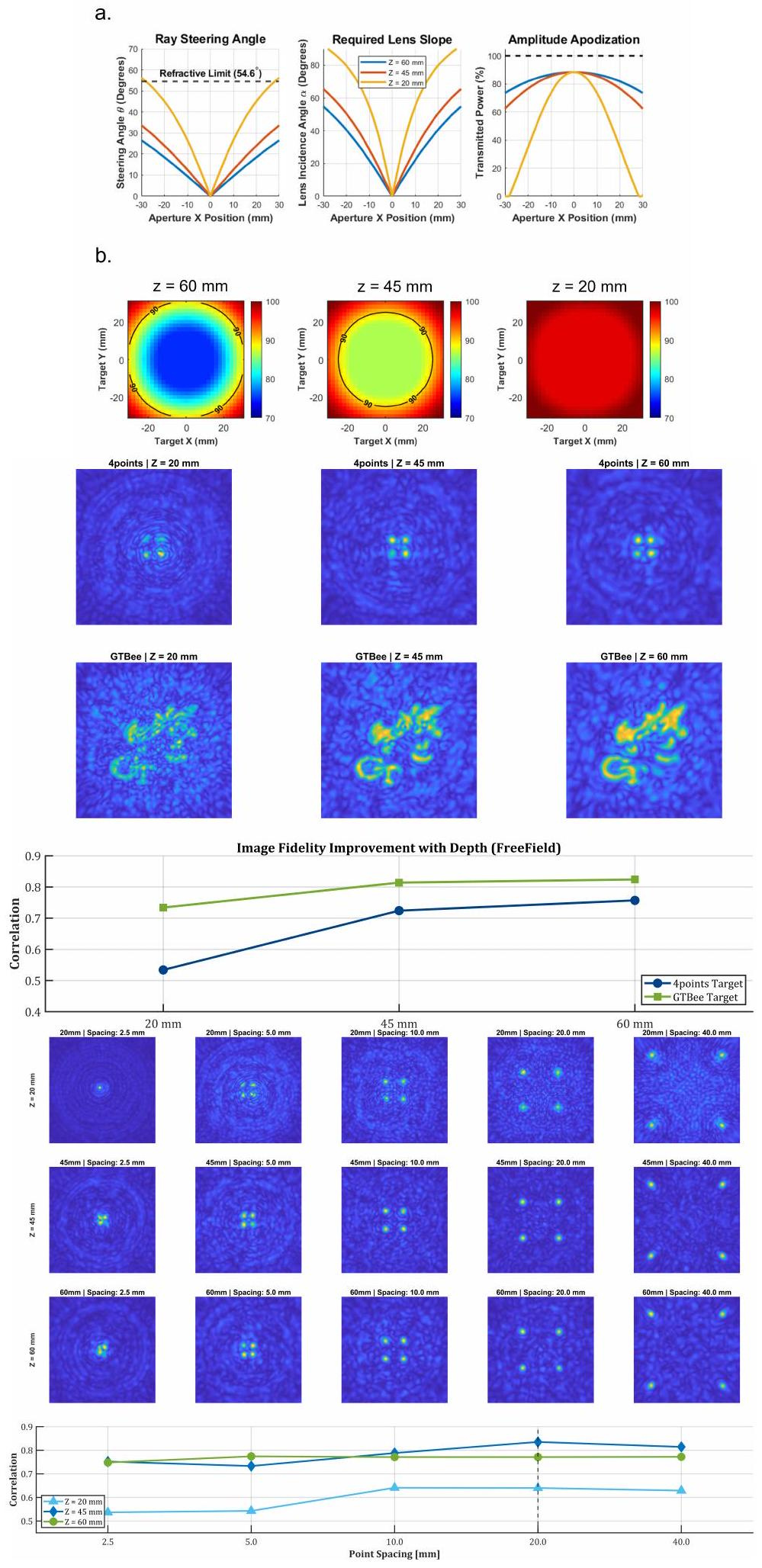}
\caption{\textbf{The high-NA failure modes are geometric rather than algorithmic, and define a quantitative operating window for free-field design.} Two compounding mechanisms limit performance as the steering angle grows: Fresnel amplitude apodization, which the one-way propagator underestimates because it imposes $R = 0$, and shear-wave mode conversion at the lens-water interface once the incidence angle exceeds the critical angle $\theta_\mathrm{c} \approx 30^\circ$. (\textit{Top}) Decreasing the focal depth from $z = 60$ mm into the extreme near field at $z = 20$ mm drives steering toward the refractive limit ($\theta_\mathrm{max} \approx 54.6^\circ$) and pushes a large fraction of the aperture beyond $\theta_\mathrm{c}$. (\textit{Middle}, \textit{bottom}) Reconstruction fidelity, as correlation and PSNR, across focal depths and lateral feature spacings, showing a sharp recovery as the focal plane moves outward. Deeper focal planes, $z \geq 45$ mm at this aperture, simultaneously recover compressional fidelity and keep most of the aperture below the shear critical angle.}
\label{fig:S12}
\end{figure}

To bound the operating window of the method we isolated its free-field failure modes, both driven by steep topographical gradients: compressional amplitude apodization and shear mode conversion (Fig.~\ref{fig:S12}). We evaluated reconstruction fidelity across focal depths ($z = 20$, 45 and 60 mm) and spatial resolutions. Focusing in the extreme near field at $z = 20$ mm forced the optimizer into physically unviable steering regimes and produced severe structural degradation: the required slopes approach the absolute refractive limit ($\theta_\mathrm{max} \approx 54.6^\circ$), causing Fresnel amplitude apodization that the one-way propagator underestimates because it imposes $R = 0$, while incidence angles beyond about 30$^\circ$ partition substantial energy into unmodelled shear waves. Moving the focal plane to 45 and 60 mm relaxed the steering angles, largely avoiding both limits, and recovered correlation and PSNR sharply for complex target patterns.

\begin{figure}[htbp]
\centering
\includegraphics[width=4.3in]{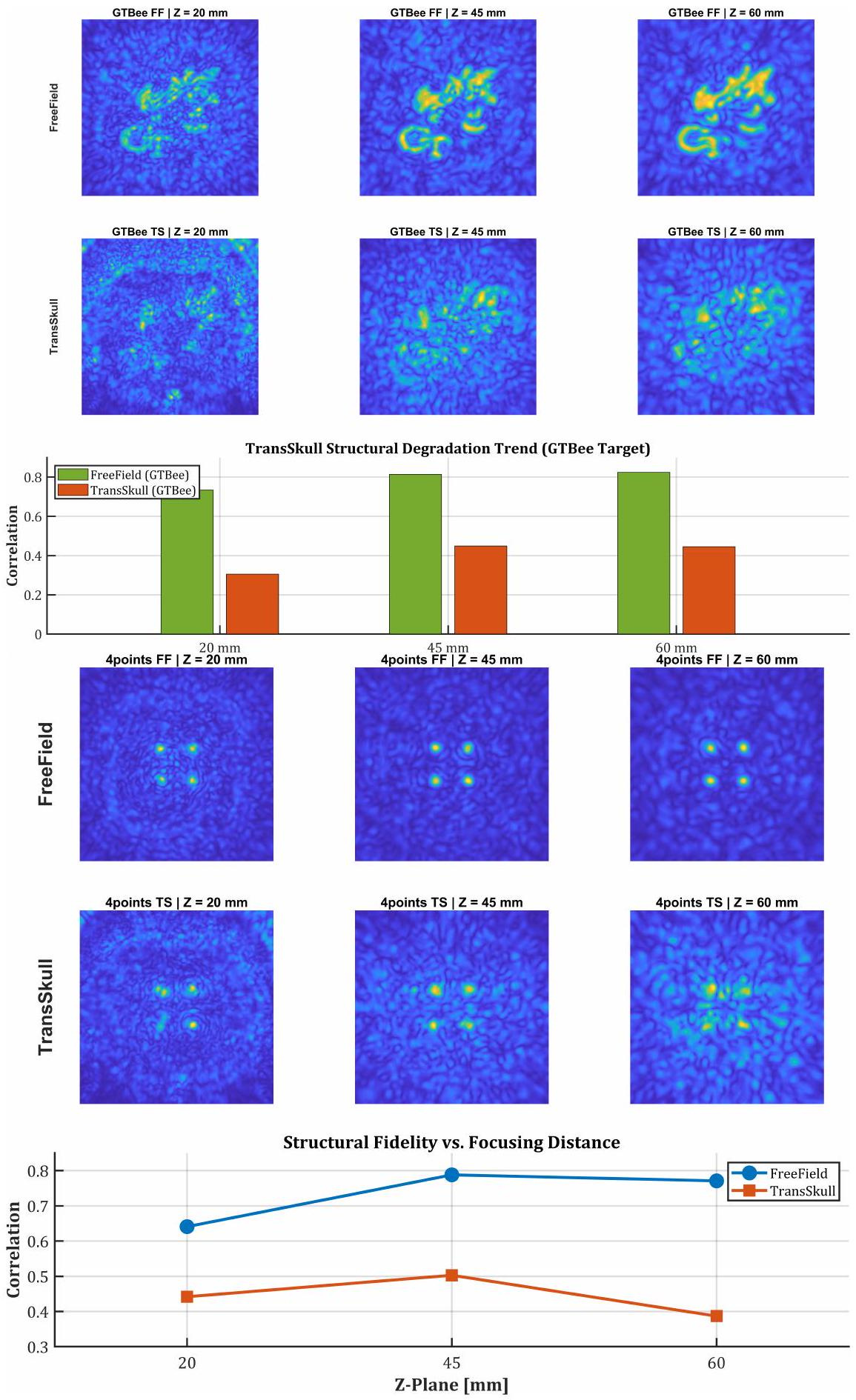}
\caption{\textbf{The boundary where fluid-based propagation fails: high-spatial-frequency, multi-target patterns through an aberrating skull push the one-way model past its validity envelope.} (\textit{Top}) GT Bee, a continuous high-spatial-frequency pattern. (\textit{Bottom}) Four-point array at three focal depths ($z = 20$, 45 and 60 mm). Free-field reconstruction (FF) is faithful; trans-skull reconstruction (TS) degrades sharply, with depth-dependent deterioration. The 45 mm plane retains partial structure, while the 20 mm near field stacks all three failure modes. These regressions trace to specific optimizer assumptions: the slowly varying envelope approximation requires the sound speed to vary smoothly over a wavelength, which is violated at the sub-millimeter water-skull discontinuity; the parabolic one-way approximation suppresses the roughly 32\% of incident energy reflected at the skull; and high incidence angles trigger unmodelled shear conversion.}
\label{fig:S13}
\end{figure}

The same boundary appears from the other side when complex patterns are projected through an aberrating skull (Fig.~\ref{fig:S13}). Free-field reconstruction is faithful, whereas trans-skull reconstruction degrades sharply and in a depth-dependent way: the 45 mm plane retains partial structure, while the 20 mm near field stacks all three failure modes, namely steep steering, severe phase aberration and near-field diffraction. These regressions trace to the specific assumptions set out above, and together they demarcate the regime in which elastic, two-way propagation would be required for further gains in fidelity.

\section{Experimental equipment}

\begin{figure}[htbp]
\centering
\includegraphics[width=3.4in]{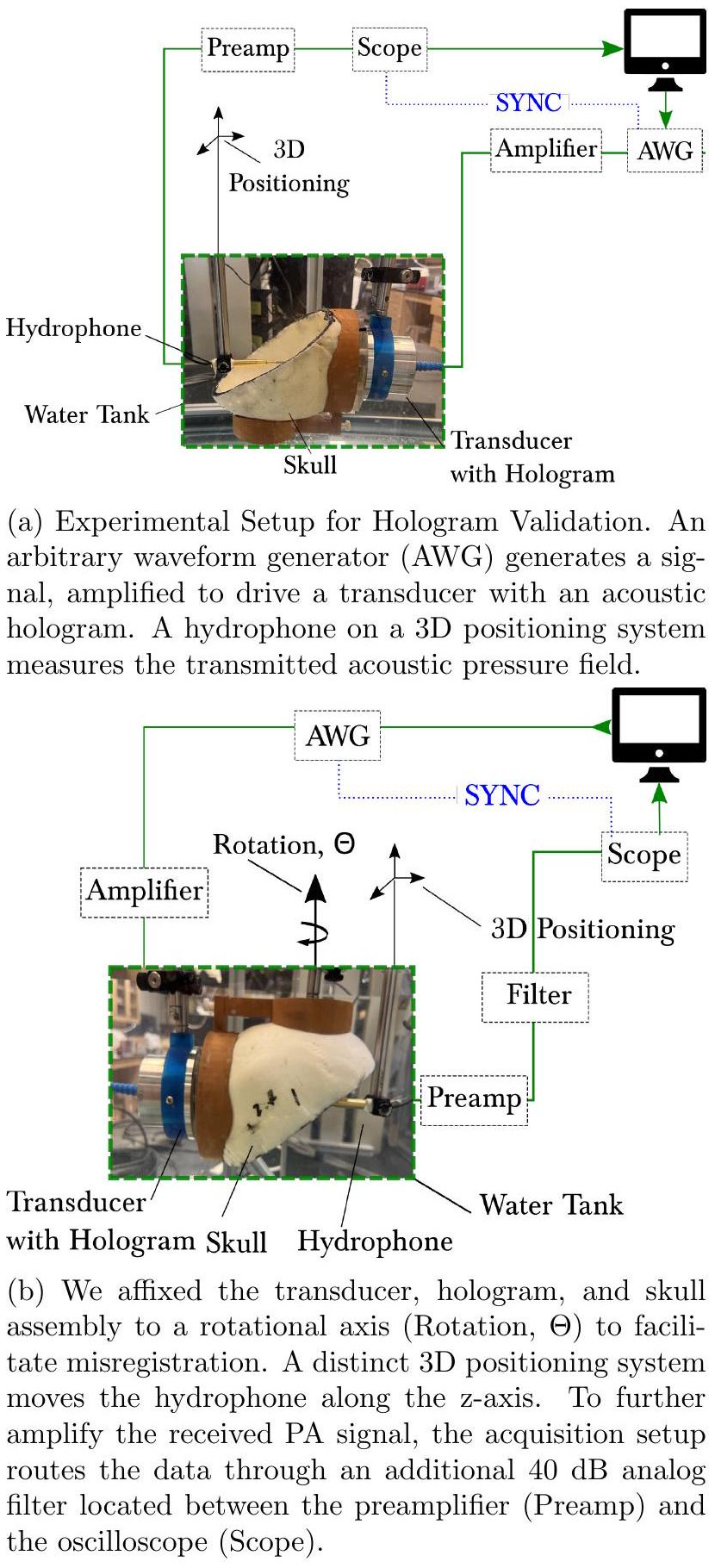}
\caption{\textbf{Experimental setups used in this study.} (\textit{a}) Setup for hologram validation. (\textit{b}) Setup for acoustic holography registration.}
\label{fig:S14}
\end{figure}

Figure~\ref{fig:S14} shows the two bench configurations used in this study: the free-field setup for hologram validation and the setup used for acoustic holography registration through the skull.

\section{Legends for Movies S1 and S2}

\movie{\textbf{Movie S1. Effect of frequency variation.} Sweeping the operating frequency from 5.0 MHz down to 0.5 MHz demonstrates the breakdown of the thin-element approximation. At lower frequencies the required lens thickness scales up as $1/f$ and the lens becomes physically bulky. That thickness maximizes the lateral walk-off distance, which severs the correspondence between entrance and exit pixels.}

\movie{\textbf{Movie S2. Effect of focusing strength.} Sweeping the focal distance from 60 mm down to 30 mm demonstrates that tighter focusing demands steeper phase gradients on the lens surface. Steeper gradients imply larger internal refraction angles $\theta_{r}$, and because the lateral walk-off scales as $\Delta x \approx L\tan\theta_{r}$, tighter focusing induces large pixel migration and grid distortion.}

\end{document}